\documentclass[pre,twocolumn,noshowpacs,amsfonts,amssymb,amsmath,eqsecnum,letterpaper,floatfix]{revtex4}

\usepackage{graphicx}  
\begin{document}   
\newlength{\GraphicsWidth}
\setlength{\GraphicsWidth}{8cm}     

\newcommand\comment[1]{\textsc{{#1}}}
\newcommand{\Label}[1]{\label{#1}}  

\renewcommand{\r}{\mathbf{r}}
\newcommand{\Man}{{\text{Manning}}}
\newcommand{\hr}{{\tilde{r}}}
\newcommand{\ha}{{\tilde{a}}}
\newcommand{\Imagin}{\Im m}
\newcommand{\eff}{{\text{eff}}}
\newcommand{\sat}{{\text{sat}}}

\title{%
Exact asymptotic 
expansions for the cylindrical Poisson-Boltzmann equation
}
\author{Gabriel T\'ellez}
\email{gtellez@uniandes.edu.co}
\affiliation{Departamento de F\'{\i}sica, Universidad de Los Andes,
A.A.~4976, Bogot\'a, Colombia}
\author{Emmanuel Trizac}
\email{trizac@lptms.u-psud.fr}
\affiliation{CNRS; Univ. Paris Sud, UMR 8626, LPTMS, Orsay Cedex, F-91405, France}
\affiliation{Center for Theoretical Biological Physics, UC San Diego,                    
       9500 Gilman Drive MC 0374 - La Jolla, CA  92093-0374, USA}

\begin{abstract}                              
The mathematical theory of integrable Painlev\'e/Toda type systems 
sheds new light on the behavior of solutions to the Poisson-Boltzmann
equation for the potential due to a long rod-like macroion.
We investigate here the case of symmetric electrolytes together with that
of 1:2 and 2:1 salts. Short and large scale features are analyzed, with a 
particular emphasis on the low salinity regime. Analytical 
expansions are derived for several quantities relevant for polyelectrolytes
theory, such as the Manning radius. 
In addition, accurate and practical expressions
are worked out for the electrostatic potential, which 
improve upon previous work and cover the full
range of radial distances. 
\end{abstract}


\maketitle


\section{Introduction}

Polyelectrolytes are polymer molecules bearing charged units.
Within the associated relevant cylindrical geometry, the long range character
of Coulombic interactions is responsible for the phenomenon
of counterion condensation. In essence, the electric potential 
created by the charged polyion features a logarithmic dependence
on radial distance. The competing entropy of confinement is of a similar
functional form, which may result in a condensation of counterions
onto the charged cylinder, when Coulombic interactions prevail, that is
when the polyelectrolyte line charge
exceeds a given threshold (see e.g. \cite{Kh}). Strongly charged 
linear polyelectrolytes thereby effectively reduce their line charge
density. This was
realized by Onsager in the 1960s and formalized 
subsequently \cite{Manning,Oosawa}. It turns out that the mean-field 
Poisson-Boltzmann (PB) theory offers a valuable framework to discuss
and analyze the phenomenon in detail 
\cite{Mora,Fixman,Weis,Ramanathan,Lebret,McCaskill,Beyerlein,Tracy-Widom-PB,Levinbis,Deserno,PLHansen,OS}.
Unfortunately, even within mean-field, exact results are scarce,
and one has to resort to numerical resolution or propose approximations.

This paper is dedicated to the derivation of exact results
within PB theory. 
We will be interested in the behavior of a unique infinite charged cylinder
of radius $a$ and uniform line charge density $e/b $, where $e>0$ is
the elementary charge and $b$ may be viewed as an equivalent average
spacing between charges along the polyion. 
This macroion is immersed in an infinite electrolyte
containing two species of microions (salt): their bulk densities $n_1$ and $n_2$ 
far from the cylinder define the Debye length $1/\kappa$ through 
$\kappa^2 = 4 \pi \ell_B (n_1 z_1^2+ n_2 z_2^2)$. Here,
$z_i e$ denotes the charge of species $i$ and $\ell_B = \beta e^2/\varepsilon$
is the Bjerrum length, which involves the solvent dielectric permittivity 
$\varepsilon$ and
the inverse temperature $\beta$.
We will focus on the three situations where analytical 
progress is possible: 1:1 electrolytes (or more generally 
symmetric $z$:$z$ electrolytes),
together with the 1:2 and 2:1 cases. 
We denote by 1:2 the situation where coions are monovalent ($z_+=1$
for a positively charged polyion) and
counterions are divalent ($z_-=-2$ or more generally, $z_- =-2 z_+$). 
On the other hand, the notation 2:1 refers 
to divalent coions with monovalent counterions. To be specific, 
introducing the dimensionless radial distance
$\hr=\kappa r$, the equations to be solved for the 
dimensionless electrostatic potential $y=\beta e \psi$ ($\psi$ being
the electric potential) read
\begin{subequations}
\Label{eq:PB}
\begin{eqnarray}
\Label{eq:PB11}
\frac{d^2 y}{d\hr^2}+\frac{1}{\hr}\frac{dy}{d\hr}=&
\sinh(y)&\qquad (1:1)\\
\Label{eq:PB12}
\frac{d^2 y}{d\hr^2}+\frac{1}{\hr}\frac{dy}{d\hr}=&
\frac{1}{3}\left(e^{2y}-e^{-y}\right)&\qquad(1:2)\\
\Label{eq:PB21}
\frac{d^2 y}{d\hr^2}+\frac{1}{\hr}\frac{dy}{d\hr}=&
\frac{1}{3}\left(e^{y}-e^{-2y}\right)&\qquad(2:1)
\end{eqnarray}
\end{subequations}
The solution should satisfy the boundary conditions 
at $\ha = \kappa a$
\begin{equation}
\Label{eq:BC}
\lim_{\hr\to\ha} \,\hr\, \frac{dy}{d\hr} \,=\, -2 \,\xi
\end{equation}
and $\lim_{\hr\to\infty}y(\hr)=0$. We have introduced here the 
dimensionless bare line charge $\xi= \ell_B/b$, that plays a key role
below. It is understood that the cylinder is positively charged,
without loss of generality \footnote{
Note that the $1:2$ case with $\xi>0$, with coions of valency $z_1=+1$ 
and counterions of valency
$z_2=-2$, is identical to the case $z_1=+2$,
$z_2=-1$ with $\xi<0$. Equation~(\ref{eq:PB21}) and the boundary
condition~(\ref{eq:BC}) is of course 
formally equivalent to Eq.~(\ref{eq:PB12})
after the substitutions $y\to-y$ and $\xi\to-\xi$.
}.

We shall not comment here the well documented 
limitations of the mean-field
approximation (see e.g. \cite{Belloni,Hansen,Grosberg,Levin,Netz,AA}),
but only summarize the main findings.
In a solvent like water at room temperature, mean-field
provides an accurate description in the 1:1 case for all existing
polyelectrolytes. We also expect the 2:1 situation to be 
correctly described, while in the 1:2 case, micro-ionic
correlations, discarded within mean-field, play
an important role at large values of $\xi$.
Schematically, the coupling parameter $\Gamma = z^{3/2} \sqrt{\xi \ell_B/(2 \pi a)}$ allows to appreciate the importance of those correlations
where $z$ stands for the counter-ions valency.
When $\Gamma<2$, mean-field holds.

The crux of our approach lies in the mapping between equations 
(\ref{eq:PB}) and Painlev\'e/Toda type equations (see below), 
that are in particular
relevant to describe the spin-spin correlator in the two-dimensional
Ising model, and have been the subject of deep mathematical 
investigations 
\cite{McCoy-Tracy-Wu,Widom-solut-Toda,TracyWidom-Toda-asympt}. 
It seems that this body
of work has not been fully transposed in the language of polyelectrolyte
physics, with the exception of the limit
$\kappa a \to 0^+$ (e.g. realized with a charged line
--of vanishing thickness--
at finite salt) 
where it allowed to show rigorously the existence
of the condensation \cite{McCaskill,Beyerlein}
and also provided exact results for the electric potential
\cite{McCaskill,Tracy-Widom-PB}. However, it will appear below
that the limit $\kappa a \to 0^+$ is approached logarithmically
slowly so that finite salt correction cannot be neglected in practice. 
It also turns that to the best of our knowledge, the case of
2:1 electrolytes has not been considered in the physics literature, 
whereas some results have been reported in the reverse 1:2 case,
but below the condensation threshold and again in the limit
$\ha \to 0$ \cite{Tracy-Widom-PB}. Our aim is to fill these gaps. 
Our formulas will prove
extremely accurate when compared to the numerical solutions
of (\ref{eq:PB}) .

The paper is organized as follows. In section \ref{sec:formal}, the
formal solutions of Eqs. (\ref{eq:PB}) will be given. As such, these
relations are not useful in practice, and deriving ``ready-to-use''
and operational expressions will be the purpose of the subsequent analysis.
Section \ref{sec:asymptotic} will then be devoted to the derivation
of asymptotic expansions for the electric potential, and to the connection 
between short and large distance features. These results identify 
a change in the short distance behavior when increasing the charge density
above a threshold $\xi_c$, what will be worked out. 
This is the fingerprint of counterion 
condensation, which turns out to be smoothed by salt
(strictly speaking, the phenomenon is critical in the limit 
$\kappa a \to 0$ only).
Section \ref{sec:limiting}
will analyze a few limiting cases of particular interest. 
Analytical expressions for several important quantities such as
the Manning radius will be given and compared in section 
\ref{sec:discussion} to the numerical data obtained 
solving the non-linear PB equation. A similar analysis will be
performed for the electric potential, which will assess the
validity of the analytical results. 
Although particular attention will be paid to the low salinity
regime (the results of sections \ref{sec:asymptotic} to \ref{sec:discussion}
rely on expansions that typically break down for $\kappa a>1$), 
we will also report in section \ref{sec:upperbound}
some results concerning the effective charge of the polyion,
valid for all values
of $\kappa a$.
A preliminary account of parts of this work has been published in 
\cite{TT}.

Before embarking on the analysis of the properties of Eqs. 
(\ref{eq:PB}), we note that for a macroion with a small
reduced linear charge density $\xi_\eff$, the problem may be linearized 
with solution
\begin{equation}
\Label{eq:DH-solut}
y_{\text{lin}}(\hr)=\frac{2 \xi_{\text{eff}}}{\ha K_1(\ha)}
K_0(\hr).
\end{equation}
Here $K_0$ and $K_1$ are the modified Bessel functions.  Since the
solution to the non-linear problem (\ref{eq:PB}) vanishes when $\hr\to\infty$ we 
necessarily have that for large distances $\hr$, $y(\hr)$ behaves
as~(\ref{eq:DH-solut}). However in the non-linear case the constant
$\xi_\eff$ is not the bare charge of the
rod. This quantity is called the effective charge and depends
on salinity conditions, bare charge, temperature etc.
In the weak overlap approximation where the double--layers
of different macroions are well separated (it is typically the
case when their mutual distance is larger than $1/\kappa$),
the (squared) effective charge governs the amplitude of 
the interaction free energy.

\section{Formal solution to the problem}
\Label{sec:formal}

Poisson--Boltzmann equation~(\ref{eq:PB11}) in the 1:1 case has been
solved in Ref.~\cite{McCoy-Tracy-Wu} in the context of the Painlev\'e
III theory, since $e^{y/2}$ obeys a particular case of Painlev\'e III
equation (see e.g. \cite{Davis} and appendix \ref{app:A}; more details
together with a relevant bibliography may be found in \cite{Benham}). 
More generally, Eqs.~(\ref{eq:PB11}), (\ref{eq:PB12})
and~(\ref{eq:PB21}) are the first equations of the hierarchy of
cylindrical Toda equations. In Ref.~\cite{Widom-solut-Toda}, a class of
solutions to this Toda equations has been reported, hence the 
solutions to the
Poisson--Boltzmann equations~(\ref{eq:PB11}), (\ref{eq:PB12}) and
(\ref{eq:PB21}).  These solutions obey a boundary condition at
$\hr\to\infty$, given in terms of a constant $\lambda$ by
\begin{subequations}
\Label{eqs:large-dist}
\begin{equation}
y_{11}(\hr)\sim 4 \lambda K_0(\hr)
\end{equation}
for the 1:1 case,
\begin{equation}
y(\hr)\sim 6 \lambda K_0(\hr)
\end{equation}
for both 1:2 and 2:1 cases.
\end{subequations}

The solutions are expressed as determinants of certain operators. For 
1:1 salts, we have
\begin{equation}
\Label{eq:sol11}
y_{11}(\hr)=2\ln\det(1+\lambda K_{\hr})-2\ln\det(1-\lambda K_{\hr})
\end{equation}
with $K_{\hr}$ an integral operator on $\mathbb{R}^{+}$ with kernel
\begin{equation}
\Label{eq:opK11}
K_{\hr}(u,v)=\frac{
e^{-\hr(u+u^{-1})/2}}{u+v}
\,.
\end{equation}
In the 1:2 case the solution reads
\begin{equation}
\Label{eq:sol12}
y_{12}(\hr)=\ln\det(1-\lambda K_{\hr}^{(0)})-\ln\det(1-\lambda
K_{\hr}^{(2)})
\end{equation}
while in the 2:1 situation it is
\begin{equation}
\Label{eq:sol21}
y_{21}(\hr)=\ln\det(1+\lambda K_{\hr}^{(2)})-\ln\det(1+\lambda
K_{\hr}^{(1)})
\end{equation}
where 
\begin{subequations}
\Label{eq:opK12-21}
\begin{eqnarray}
K_{\hr}^{(0)}&=&(\zeta-\zeta^2)O^{(1)}_{\hr}+(\zeta^2-\zeta)O^{(2)}_{\hr}\\
K_{\hr}^{(1)}&=&(\zeta^2-1)O^{(1)}_{\hr}+(\zeta-1)O^{(2)}_{\hr}\\
K_{\hr}^{(2)}&=&(1-\zeta)O^{(1)}_{\hr}+(1-\zeta^2)O^{(2)}_{\hr}.
\end{eqnarray}
\end{subequations}
Here $\zeta=e^{i 2\pi/3}$ and $O^{(1)}_{\hr}$ and
$O^{(2)}_{\hr}$ are integral operators on $\mathbb{R}^{+}$ with kernel
\begin{equation}
O_{\hr}^{(1)}(u,v)=\frac{e^{-\hr[(1-\zeta)u+(1-\zeta^2)u^{-1}]/(2\sqrt{3})
}}{-\zeta u +v}
\end{equation}
and $O_{\hr}^{(2)}(u,v)=\overline{O_{\hr}^{(1)}(u,v)}$, the bar
denoting complex conjugation. It can be shown~\cite{Widom-solut-Toda}
that $\det(1-\lambda K_{\hr}^{(1)})=\det(1-\lambda K_{\hr}^{(0)})$
(making use of a change of variable $u\to u^{-1}$), thus from the
solution to the case 1:2, $y_{12}$, one can obtain the solution for
the case 2:1 as $y_{21}=-y_{12}$ with the change
$\lambda\to-\lambda$, as previously announced.

To solve completely the problem we are interested in, we should impose
the boundary condition~(\ref{eq:BC}) to express $\lambda$ in terms
of the bare charge $\xi$. Notice that the constant $\lambda$
introduced above is closely related to the effective charge
$\xi_\eff$. Indeed, 
\begin{subequations}
\Label{eq:xieff-lambda}
\begin{eqnarray}
\Label{eq:xieff-lambda11}
\xi_\eff=&2 \ha K_1(\ha)\, \lambda \qquad (1:1)\\
\xi_\eff=&3 \ha K_1(\ha)\, \lambda \qquad (1:2)\\
\xi_\eff=&3 \ha K_1(\ha)\, \lambda \qquad (2:1)
\end{eqnarray}
\end{subequations}
The determination of the effective charge in terms of the bare one is
interesting {\it per se}, 
since it allows to use, at large distances, the
linear theory expression, provided that the bare charge is replaced by the
effective one. 
Writing down the boundary condition~(\ref{eq:BC}) gives the bare
charge as a function of the effective one. In the limiting
case $\ha \to 0$, this relation may be inverted
since the associated (so-called) connection problem was solved: knowing
the large-distance behavior~(\ref{eqs:large-dist}) of the
solutions~(\ref{eq:sol11}), (\ref{eq:sol12}) and~(\ref{eq:sol21}),
the short-distance behavior for $\hr\to0$ 
follows~\cite{TracyWidom-Toda-asympt}. The 1:1 and 1:2 cases for $a=0$
were studied in detail by Tracy and Widom in
Ref.~\cite{Tracy-Widom-PB}. The 2:1 situation, even for $a=0$, was not
explicitly considered in Ref.~\cite{Tracy-Widom-PB}, although it can
be obtained following the same lines as those exposed
in~\cite{Tracy-Widom-PB}. 

The situation with $\ha \neq 0$ is analytically more difficult,
but we will show below that useful expressions may nevertheless
be derived there, that rely on accurate approximations for certain
key quantities related to the parameter $\lambda$.
These expressions allow to fully cover the regime of 
thin cylinders and/or low salt $\ha <1$, 
whereas the discussion of results valid at arbitrary salt content 
is deferred to section \ref{sec:upperbound}.

\section{Asymptotic expansions}
\Label{sec:asymptotic}

The results reported are obtained from the limit $\ha \to 0$
but turn out to be reliable for $\ha <1$ (see below).

\subsection{Short distance behavior}
\Label{ssec:short}
The general results of Ref.~\cite{TracyWidom-Toda-asympt} for the cylindrical
Bullough-Dodd equation (which belongs to the Toda family and onto which 
the above asymmetric PB equation can be mapped) 
allow to find the 
small distance asymptotics of $y(\hr)$ given in Eq. (\ref{eq:sol12})
for 1:2 electrolytes.
Allowing for negative values of $\xi$ will therefore
also provide the solution to the 2:1 case. Similarly, the 1:1 
behavior encoded in
Eq.~(\ref{eq:sol11})
follows from Theorem 3 of Ref. \cite{McCoy-Tracy-Wu}
and its corollaries (in particular those derived in section
IV.I of \cite{McCoy-Tracy-Wu}).

It turns that the parameter $\lambda$ plays an essential role here.
Its position with respect to a threshold value $\lambda_c$
discriminates two different short scale behaviors.
The threshold values read:
\begin{subequations}
\Label{eq:lambdac}
\begin{eqnarray}
\lambda_c^{(1:1)}&=&\frac{1}{\pi} \\
\lambda_c^{(1:2)}&=&\frac{1}{2\sqrt{3}\,\pi}  \\
\lambda_c^{(2:1)}&=&\frac{\sqrt{3}}{2\,\pi}  .
\end{eqnarray}
\end{subequations}

We will show that the condition $\lambda<\lambda_c$ is equivalent to
$\xi<\xi_c$ where $\xi_c$ is some threshold charge. In the limit
$\kappa a = \ha \to 0$, $\xi_c$ coincides with the known 
Manning parameter beyond which counterion condensation sets in:
\begin{eqnarray}
\xi_{\Man}^{(1:1)} &=& 1 \\
\xi_{\Man}^{(1:2)} &=& 1/2\\
\xi_{\Man}^{(2:1)} &=& 1.
\end{eqnarray}
However, as soon as $\ha \neq 0$, it will appear that 
$\xi_c$ significantly differs from $\xi_{\Man}$.
For $\lambda<\lambda_c$, we have \cite{McCoy-Tracy-Wu,McCaskill}
\begin{widetext}
\begin{subequations}
\Label{eq:small-asymptotics}
\begin{eqnarray}
\Label{eq:small-asymptotics11}
y_{11}(\hr)=&\displaystyle
-2A\ln\hr+2\ln B 
-2\ln\left[1-\frac{B^2\hr^{2-2A}}{16(1-A)^2}
    \right]+{\cal O}(\hr^{2+2A})
&\qquad(1:1)\\
\Label{eq:small-asymptotics12}
y_{12}(\hr)=&\displaystyle 
-2A \ln\hr +\ln B 
-\ln\left[1-\frac{B^2 \hr^{2-4A}}{12(2A-1)^2}\right]+
{\cal O}(\hr^{2+2A})
&\qquad(1:2)\\
\Label{eq:small-asymptotics21}
y_{21}(\hr)=&\displaystyle 
-2A \ln\hr +\ln B -
2\ln\left[
1-\frac{\hr^{2-2A} B}{24(1-A)^2}
\right]
+{\cal O}(\hr^{2+4A})
&\qquad(2:1)
\end{eqnarray}
\end{subequations}
where $A$ is a function of $\lambda$, see
Eqs.~(\ref{eq:connection11a}) and (\ref{eq:Alambda}) below, and
\begin{subequations}
\label{eq:B}
\begin{eqnarray}
B=&
\displaystyle
2^{3A}\ \ 
\frac{\Gamma\left(\frac{1+A}{2}\right)}{
\Gamma\left(\frac{1-A}{2}\right)}
&\qquad(1:1)\\
\Label{eq:B12}
B=&
\displaystyle
3^{3A}2^{2A}\frac{\Gamma\left(\frac{1+A}{3}\right)
\Gamma\left(\frac{2+2A}{3}\right)}{\Gamma\left(\frac{2-A}{3}\right)
\Gamma\left(\frac{1-2A}{3}\right)}
&\qquad(1:2)\\
\Label{eq:B21}
B=&
\displaystyle
3^{3A}2^{2A}\frac{\Gamma\left(\frac{2+A}{3}\right)
\Gamma\left(\frac{1+2A}{3}\right)}{\Gamma\left(\frac{1-A}{3}\right)
\Gamma\left(\frac{2-2A}{3}\right)}
&\qquad(2:1)
\end{eqnarray}
\end{subequations}
\end{widetext}
Here $\Gamma$ is the Euler gamma function.

On the other hand, for $\lambda >\lambda_c$ (or equivalently 
$\xi>\xi_c$), the short distance behavior is characterized by a
parameter $\mu>0$ and
we have
\begin{subequations}
\Label{eq:pttedistgrdlambda}
\begin{eqnarray}
e^{-y_{11}/2} &\simeq& \frac{\hr}{4 \mu} \,\sin\left[
-2 \mu \log(\hr) - 2 \mu\,{\cal C}^{(1:1)}
\right]
\Label{eq:pttedistgrdlambda11}\\
e^{-y_{12}} &\simeq& \frac{\hr}{3 \sqrt{3}\,\mu} \,\sin\left[
-3 \mu \log(\hr) - 3 \mu\,{\cal C}^{(1:2)}
\right]
\Label{eq:pttedistgrdlambda12}\\
e^{-y_{21}/2} &\simeq& \frac{\sqrt{2} \hr}{3 \sqrt{3}\,\mu} \,\sin\left[
-\frac{3}{2} \mu \log(\hr) - \frac{3}{2} \mu\,{\cal C}^{(2:1)}
\right].
\Label{eq:pttedistgrdlambda21}
\end{eqnarray}
\end{subequations}
For the sake of notational simplicity, we do not explicitly mention that 
$\mu$ depends on the situation 1:1, 1:2 or 2:1 considered. The constants
${\cal C}$ appearing above are defined as
\begin{eqnarray}
\Label{eq:C}
{\cal C}^{(1:1)} &=& \gamma - 3 \log 2 \,\simeq\, -1.502
\\
{\cal C}^{(1:2)} &=& \gamma - \frac{1}{3}\log 2- \frac{3}{2}\log 3 
\,\simeq\, -1.301
\\
{\cal C}^{(2:1)} &=& \gamma - \log 2- \frac{3}{2} \log 3 
\,\simeq\, -1.763
\end{eqnarray}
where $\gamma\simeq 0.5772...$ is the Euler constant.  We emphasize
that expressions (\ref{eq:pttedistgrdlambda}) do not have the same
status of mathematical rigor as their counterparts
(\ref{eq:small-asymptotics}) valid for $\lambda<\lambda_c$, as appears
in Appendix \ref{app:B}. The constants ${\cal C}$ in
(\ref{eq:pttedistgrdlambda}) arise from the linearization of a
$\mu$-dependent function, which is justified from a practical point of
view since for physically relevant situations, $\mu$ is small enough
to allow for the corresponding Taylor expansions.  For the exact
expressions see Appendix \ref{app:B}.

We note here that the relation between surface potential and 
surface charge follows from the expressions given in the present
section.
At this point, the electric potential depends on a parameter 
$A$ for $\lambda<\lambda_c$ or on a parameter $\mu$ for $\lambda >\lambda_c$.
These two parameters are related to the bare reduced charge $\xi$
of the cylindrical polyion through the boundary condition
(\ref{eq:BC}). Before clarifying this relation, we precise
the connection between small $\hr$ and large $\hr$ behavior.

\subsection{Connecting short and large scale features}
\Label{ssec:connect}
For 1:1 electrolytes, the connection problem was solved in 
\cite{McCoy-Tracy-Wu,McCaskill}, with the result
\begin{eqnarray}
&\displaystyle\lambda = \frac{1}{\pi} \, \sin\left(\frac{\pi A}{2}\right)
&\quad \hbox{for} \quad \lambda < \pi^{-1} \,(\text{or } \xi<\xi_c)\ \
  \ 
\Label{eq:connection11a}\\
&\displaystyle\lambda = \frac{1}{\pi}\, \cosh(\pi \mu)
&\quad \hbox{for} \quad \lambda > \pi^{-1} \,(\hbox{or } \xi>\xi_c)
\Label{eq:connection11b}
\end{eqnarray}
Since $\lambda$ is related to the effective charge with governs the 
far field behavior [see (\ref{eq:xieff-lambda})], the above 
expressions realize an interesting connection between small
and large $\hr$ features.

For asymmetric electrolytes and $\lambda<\lambda_c$
\begin{subequations}
\Label{eq:Alambda}
\begin{eqnarray}
A=&
\displaystyle
-\frac{1}{4}+\frac{3}{2\pi}\arcsin\left(\frac{1}{2}+
\frac{\lambda}{2\lambda_c}\right)
&\quad(1:2)~~
\\
\Label{eq:A12}
A=&
\displaystyle
\frac{1}{4}+\frac{3}{2\pi}\arcsin\left(-\frac{1}{2}+
\frac{3\lambda}{2\lambda_c}\right)
&\quad(2:1)~~
\end{eqnarray}
\end{subequations}
In these two expressions, the values of $\lambda_c$ differ
[$\lambda_c^{(2:1)} = 3 \lambda_c^{(1:2)}$ see (\ref{eq:lambdac})]
in such a way that expressing $A$ as a function of $\lambda_c^{(1:2)}$
in both cases would provide the same expression, but a change in sign
in $A$ and in $\lambda$. This reflects the original symmetry of 
1:2 and 2:1 situations, when $\xi$ is allowed to change sign. 
This symmetry is broken here by the choice $\xi>0$ in both
cases, which illustrates the practical difference between 
them.

From a more formal point of view, the
expressions~(\ref{eq:connection11a}) and~(\ref{eq:Alambda}) hold even
for complex parameters, provided by
\begin{subequations}
\Label{eq:lambda-conditions}
\begin{eqnarray}
\lambda&\notin&
(-\infty,-\lambda_c^{(1:1)}]\cup[\lambda_c^{(1:1)},\infty)
\quad (1:1)
\\
\lambda&\notin&
(-\infty,-\lambda_c^{(2:1)}]\cup[\lambda_c^{(1:2)},\infty)
\quad (1:2)
\\
\lambda&\notin&
(-\infty,-\lambda_c^{(1:2)}]\cup[\lambda_c^{(2:1)},\infty)
\quad (2:1)
\end{eqnarray}
\end{subequations}

Alternatively, for $\lambda>\lambda_c$, we have
\begin{subequations}
\Label{eq:mulambda}
\begin{eqnarray}
\frac{\lambda}{\lambda_c}+1&=& 2\cosh(\pi \mu)
\qquad(1:2)
\\
\frac{3\lambda}{\lambda_c}-1&=& 2\cosh(\pi \mu)
\qquad(2:1)
\end{eqnarray} 
\end{subequations}

\subsection{Applying the boundary condition at polyion contact}
\Label{ssec:bc}
In order to have a closed problem, we need to impose the boundary condition
(\ref{eq:BC})
at $\hr= \ha$ which relates $A$ to $\xi$ (and therefore $\lambda$ to
$\xi$ from the connection formulae) if $\lambda<\lambda_c$ and
similarly relates $\mu$ to $\xi$ if $\lambda>\lambda_c$.

A straightforward computation gives in the 1:1 situation
\begin{eqnarray}
&&\xi \,=\, A - \frac{(2-2A) (\kappa a)^{2-2 A}}{16 (1-A)^2
B^2-(\kappa a)^{2-2 A}}
\Label{eq:bc11a}\\
&& (\xi-1) \tan\left[
2 \mu \log(\kappa a/8) +2 \mu \gamma
\right] \,=\, 2 \mu
\Label{eq:bc11b}
\end{eqnarray}
where it is understood that the first line holds for $\lambda<\lambda_c$
while the second is relevant for $\lambda>\lambda_c$, in which case
$\mu$ is the smallest positive root of the equation.
Here again, $\gamma$ denotes the Euler constant.
The dependence of $\mu$ and $A$ upon $\xi$ is shown in
Figures \ref{fig:A11} and \ref{fig:mu11}.

\begin{figure}[htb]
\includegraphics[height=5.5cm,angle=0]{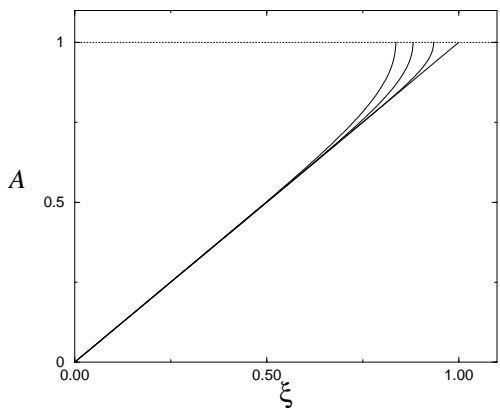}
\vskip -5mm
\caption{Short scale exponent $A$ (relevant for $\lambda<\lambda_c$) 
as a function of bare charge
for a 1:1 salt. Four salinities are displayed: from left to
right,  $\ha = 10^{-2}$,  $10^{-3}$, $10^{-6}$ and $0^+$,
the latter coinciding with the first bissectrix. The values of
$\xi$ where $A=1$ correspond to an end point which defines
the threshold bare charge $\xi_c$.}
\Label{fig:A11}
\end{figure}

\begin{figure}[htb]
\includegraphics[height=5.5cm,angle=0]{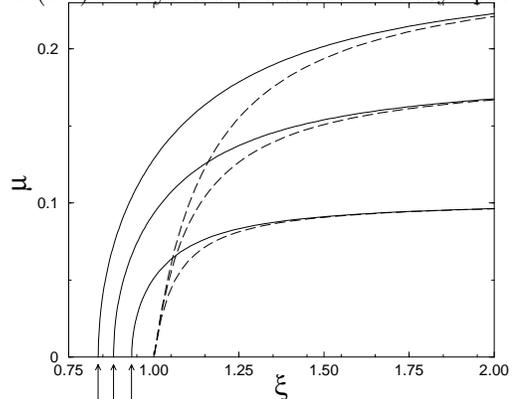}
\vskip -5mm
\caption{Short scale exponent $\mu$ relevant for $\lambda>\lambda_c=1/\pi$ versus
reduced bare charge (1:1 salt). 
The three curves correspond from left to
right to  $\ha=10^{-2}$,  $10^{-3}$ and $10^{-6}$. The limit $\ha \to 0$ gives
$\mu=0$. The salt dependent threshold values $\xi_c$ are shown 
by the arrows, which correspond to the end points where $A=1$ in Figure
\ref{fig:A11}. The dashed curves correspond to approximation 
(\ref{eq:mu11approx}).}
\Label{fig:mu11}
\end{figure}

In addition for the 1:2 situation, we get
\begin{eqnarray}
&&\xi \,=\, A - \frac{(1-2A)(\kappa a)^{2-4A}B^2}{
    12(2A-1)^2-B^2(\kappa a)^{2-4A}}
\Label{eq:bc12a}\\
&& (2\xi-1) \tan\left[
3 \mu \log(\kappa a) +3 \mu\, {\cal C}^{(1:2)}
\right] \,=\, 3 \mu~~
\Label{eq:bc12b}
\end{eqnarray}
while for the 2:1 case
\begin{eqnarray}
&&\xi \,=\, A - \frac{2(1-A)(\kappa a)^{2-2A} B}{
    24(1-A)^2-(\kappa a)^{2-2A} B}
\Label{eq:bc21a}\\
&& (\xi-1) \tan\left[
\frac{3}{2} \mu \log(\kappa a) +\frac{3}{2} \mu\, {\cal C}^{(2:1)}
\right] \,=\, \frac{3}{2}\, \mu~~
\Label{eq:bc21b}
\end{eqnarray}
The values of $B$ are given in (\ref{eq:B}).

At this point, we emphasize that the above expressions
are not mathematically exact, since use was made of the
asymptotic expansions given in section \ref{ssec:short}
to compute $dy/d\hr$ in (\ref{eq:BC}).
They nevertheless become asymptotically exact for $\kappa a=\ha \to 0$
\footnote{
In this respect, there is a further approximation in 
(\ref{eq:bc11b}) compared to (\ref{eq:bc11a}) since it 
relies on (\ref{eq:pttedistgrdlambda21}), which is not exact.
A similar remark holds for (\ref{eq:bc12b}) or (\ref{eq:bc21b}).
It is however a simple task to get rid of this extra degree
of approximation using the expressions given in Appendix \ref{app:B}.
In any case, the resulting relation between $\xi$ and $\mu$ is 
only asymptotically exact in the limit $\ha \to 0$,
and resorting to the full expressions of Appendix \ref{app:B}
brings very little improvement in terms of numerical accuracy.
},
and will be shown to provide excellent results for the full range
of thin cylinders $\ha <1$.
 
  Finally, it is important to remember that 
\begin{eqnarray}
    A \to \xi_{\Man} & \hbox{ for} & \lambda \to \lambda_c^- 
    \quad(\hbox{or } \xi \to \xi_c^-) 
    \\
    \mu \to 0 & \hbox{ for} & \lambda \to \lambda_c^+ 
    \quad (\hbox{or } \xi \to \xi_c^+).
\end{eqnarray}
In the vicinity of the threshold $\xi_c$, it can be checked that
$A-\xi_{\Man}$ and $\mu$ scale like $\sqrt{\xi-\xi_c}$, as may be
expected from Figures \ref{fig:A11} and \ref{fig:mu11}.

\subsection{What are the associated threshold potentials and charges ?}
\Label{ssec:crit}
To understand the change of behavior of the electrostatic potential
below or above $\lambda_c$, it is instructive to compute the
associated threshold charge $\xi_c$. There are several ways to 
perform such a computation, either from below taking the limit
$\lambda\to \lambda_c^-$ and considering the relations
between $A$ and $\xi$, or from above ($\lambda\to\lambda_c^+$)
manipulating $\mu$. By construction (see Appendix \ref{app:B} 
for the details of the analytic continuation method
used), these two routes provide the same short distance 
potentials, that read
for the three different electrolytes
\begin{widetext}
\begin{subequations}
\Label{eq:criticalpot}
\begin{eqnarray}
y_{11}(\hr) &\stackrel{\lambda=\lambda_c}{=}& -2 \log\left(\frac{\hr}{2}\right) 
-2 \log\left[-\log(\hr)-{\cal C}^{(1:1)}
\right]~~~~ 
\Label{eq:crit11}\\
y_{12}(\hr) &\stackrel{\lambda=\lambda_c}{=}&- \log\left(\frac{\hr}{\sqrt{3}}\right) 
- \log\left[-\log(\hr)-{\cal C}^{(1:2)}
\right]~~~~ 
\Label{eq:crit12}\\
y_{21}(\hr) &\stackrel{\lambda=\lambda_c}{=}& -2 \log\left(\frac{\hr}{\sqrt{6}}\right) 
-2 \log\left[-\log(\hr)-{\cal C}^{(2:1)}
\right]~~~~ 
\Label{eq:crit21}
\end{eqnarray}
\end{subequations}
\end{widetext}
Limiting ourselves to the region $\ha<1$ (as for all the results
reported in this section \ref{sec:asymptotic})) ensures here that the
previous expressions are real.  Equations (\ref{eq:crit11}) and
(\ref{eq:crit12}) were already given in
\cite{McCoy-Tracy-Wu,McCaskill,TracyWidom-Toda-asympt,Tracy-Widom-PB}
and are repeated here for the sake of completeness (expression
(\ref{eq:crit12}) appeared under number (3.2) with misprints in Ref.
\cite{Tracy-Widom-PB}, where the notation 21 used there corresponds to
our 1:2 case \footnote{ We also note that references
\cite{McCaskill,Beyerlein} suffer from several typographic errors that
make detailed comparison quite difficult.  }).

From Eqs. (\ref{eq:criticalpot}), it is straightforward to
compute the integrated charge $q(r)$ in a cylinder of radius
$r$ which reads $q(\hr)=-(\hr/2) dy/d\hr$ from Gauss' law. 
Evaluating this expression at $\hr=\ha$, we recover 
(\ref{eq:BC}), hence the value of $\xi$ associated to
$\lambda_c$
\begin{subequations}
\Label{eq:xicrit}
\begin{eqnarray}
\xi_c^{(1:1)} &=& 1 + \frac{1}{\log (\kappa a) + {\cal C}^{(1:1)}}
\Label{eq:xicrit11}
\\
\xi_c^{(1:2)} &=& \frac{1}{2} + \frac{1}{2 \log (\kappa a) + 2 \,{\cal C}^{(1:2)}}
\Label{eq:xicrit12}
\\
\xi_c^{(2:1)} &=& 1 + \frac{1}{\log (\kappa a) + {\cal C}^{(2:1)}} 
\Label{eq:xicrit21}
\end{eqnarray}
\end{subequations}
where the constants $\cal C$ are given in (\ref{eq:C}).
We will see in section \ref{ssec:ha0}
that these threshold values, which discriminate $\lambda<\lambda_c$
behavior from that at $\lambda>\lambda_c$, can be associated with 
the phenomenon of counterion condensation.

Expressions (\ref{eq:xicrit}) correspond to the leading order behavior
of $\xi_c$ and may be improved from the accurate expressions
obtained in \cite{McCoy-Tracy-Wu,TracyWidom-Toda-asympt}.
In the 1:1 case, inclusion of the so far omitted dominant correction
leads to 
\begin{equation}
e^{-y_{11}/2} = \frac{\hr}{2}\, \Omega(\hr) +\frac{\hr^5\,\Omega(\hr)}{2^9}
+ {\cal O}\left(\hr^5\, \log^2(\hr)
\right),
\Label{eq:betterxic}
\end{equation}
with 
\begin{equation}
\Omega(\hr) \,=\, -\log(\hr) - {\cal C}^{(1:1)}.
\end{equation}
Equation (\ref{eq:crit11}) corresponds to truncating the 
right hand side of (\ref{eq:betterxic}) after the first term.
The integrated charge $q(r)$ associated to (\ref{eq:betterxic})
reads, including only the dominant correction 
to the $q(r)$ leading to (\ref{eq:xicrit11})
\begin{equation}
q(\hr) \,=\, 1 -\frac{1}{\Omega(\hr)} + \frac{\hr^4 \,\Omega^2(\hr)}{16}
+ {\cal O}\left(\hr^3 \,\Omega^3(\hr) \right).
\Label{eq:qcritbetter}
\end{equation}
From $q(\ha)=\xi$, we get that the error made in (\ref{eq:xicrit11})
---which corresponds to the first two terms on the right hand side of
(\ref{eq:qcritbetter}) evaluated at $\hr=\ha$---
is of order $\ha^4 (\log\ha)^2$. More quantitatively, the terms
neglected in (\ref{eq:xicrit11}) are below $2\times10^{-3}$ and are therefore
irrelevant given that the term in $1/\log\ha$ in (\ref{eq:xicrit11})
may be of order 0.1 or larger under reasonable salt conditions.
A similar analysis could be performed to improve over expressions
(\ref{eq:xicrit12}) and (\ref{eq:xicrit21}).

Finally, note that exactly at $\xi=\xi_c$, effective charges 
take a particularly
simple form [making use of Eqs. (\ref{eq:xieff-lambda}) and
(\ref{eq:lambdac})] 
since then $\lambda=\lambda_c$.

\section{A few limiting cases}
\Label{sec:limiting}

\subsection{$a = 0$ and arbitrary $\xi$}
\Label{ssec:ha0}

It is instructive to discuss first the case $a=0$ (at fixed $\kappa>0$)
where the analytical solutions take simple forms and $\lambda$ can be
found explicitly. From Eqs. (\ref{eq:bc11a}), (\ref{eq:bc12a}) or
(\ref{eq:bc21a}) we find $A=\xi$ and
\begin{subequations}
\Label{eq:lambda}
\begin{eqnarray}
\lambda=&\displaystyle
\frac{1}{\pi}\sin\frac{\pi \xi}{2}
&\ (1:1)\ \ \ \ \ \\
\lambda=&\displaystyle
\frac{1}{2 \sqrt{3}\pi}\left[2\sin\left(\frac{2\pi}{3}\left(\xi+\frac{1}{4}
\right)\right)-1\right]
&\ (1:2)\\
\lambda=&\displaystyle
\frac{1}{2 \sqrt{3}\pi}\left[2\sin\left(\frac{2\pi}{3}\left(\xi-\frac{1}{4}
\right)\right)+1\right]
&\ (2:1)
\end{eqnarray}
\end{subequations}
From these results we obtain the effective charge of the 
rod. For $a = 0$, we simply have [see Eq. (\ref{eq:xieff-lambda})] 
$\xi_\eff=2\lambda$ (1:1) and
$\xi_\eff=3\lambda$ (1:2) or (2:1), thus
\begin{subequations}
\Label{eq:laeffkap0-all}
\begin{eqnarray}
\label{eq:111}
\xi_{\eff}=
&
\displaystyle
\frac{2}{\pi}\sin\frac{\pi \xi}{2}
&(1:1)
\\
\Label{eq:laeffkap012}
\xi_{\eff}=
&
\displaystyle
\frac{\sqrt{3}}{2\pi}
\left[2\,\sin\left[\frac{2\pi}{3}\left(
\xi + \frac{1}{4}\right)\right]-1\right]
&(1:2)\\
\xi_{\eff}=
&
\displaystyle
\frac{\sqrt{3}}{2\pi}
\left[2\,\sin\left[\frac{2\pi}{3}\left(
\xi - \frac{1}{4}\right)\right]+1\right]
&(2:1)
\Label{eq:laeffkap021}
\end{eqnarray}
\end{subequations}
Eqs. (\ref{eq:111}) and (\ref{eq:laeffkap012}) may be
found in \cite{McCaskill} and \cite{Tracy-Widom-PB}.
These formulas are valid only if the
conditions~(\ref{eq:lambda-conditions}) on $\lambda$ are
satisfied. Given that, for $ a = 0$, the threshold charges 
(\ref{eq:xicrit}) read
\begin{subequations}
\Label{eq:xicrit0}
\begin{eqnarray}
\xi_c^{(1:1)} &=& \xi_\Man^{(1:1)} \,=\,1 
\Label{eq:xicrit110}
\\
\xi_c^{(1:2)} &=& \xi_\Man^{(1:2)} \,=\, \frac{1}{2} 
\Label{eq:xicrit120}
\\
\xi_c^{(2:1)} &=& \xi_\Man^{(2:1)} \,=\, 1
\Label{eq:xicrit210}
\end{eqnarray}
\end{subequations}
expressions (\ref{eq:laeffkap0-all}) hold provided
\begin{subequations}
\Label{eq:Manning-crit}
\begin{eqnarray}
\xi<1&&(1:1)\\
\xi<1/2&&(1:2)\\
\xi<1&&(2:1)
\end{eqnarray}
\end{subequations}
or, in the most general case allowing for negative values of the
bare charge:
\begin{subequations}
\begin{eqnarray}
-1<&\xi<1&\qquad(1:1)\\
-1<&\xi<1/2&\qquad(1:2)\\
-1/2<&\xi<1&\qquad(2:1)
\end{eqnarray}
\end{subequations}
We recover here the Onsager-Manning-Oosawa criterion for counterion
condensation~\cite{Manning,Oosawa,McCaskill,Beyerlein,Deserno,OS,TT}. Strictly speaking, if $\xi$ does
not satisfy~(\ref{eq:Manning-crit}) there is no physical solution of
the Poisson--Boltzmann equation for $a=0$. As explained
in~\cite{Manning}, if the bare charge $\xi$ is above the
Manning-Oosawa threshold given in Eqs.~(\ref{eq:Manning-crit}), the
Boltzmann factor of the interaction between the rod and a counterion
is not integrable at short distances, and thus would lead to a
collapse between the macroion and the counterions. The situation is
similar to the one in the theory of two-dimensional Coulomb
systems~\cite{SamajTravenec,CornuJanco-CGas}, with logarithmic
interaction, where a system of point particles is stable only if the
charge of the particles is small enough so that the Boltzmann factor
of the interaction between two oppositely charged particles is
integrable. Otherwise only a system with hard-core particles (or with
other short-distance regularization of the Coulomb potential) is
stable.

For the 1:2 electrolyte, the value $\xi=\xi_{\Man}^{(1:2)}=1/2$
corresponds to Manning-Oosawa threshold~\cite{Manning,Oosawa}. Beyond
the threshold ($\xi\geq 1/2$), the effective charge $\xi_\text{eff}$
attains its saturation value, $ \xi_\text{eff}^{\text{sat}} =
\sqrt{3}/(2\pi) \simeq 0.275$.  One may check that in the linear
regime (weak charges, $\xi\ll 1$), $\xi_\text{eff}$ and
$\xi$ coincide, as they should.
For the 2:1 electrolyte, the Manning-Oosawa threshold is for
$\xi=\xi_{\Man}^{(2:1)}=1$ and beyond this threshold the effective
saturated charge is given by $\xi_{\text{eff}}^{\text{sat}}
=3\sqrt{3}/(2\pi)$; it is three times larger than in the 1:2 case.
In the symmetric 1:1 electrolyte case, the Manning-Oosawa threshold is for
$\xi=\xi_{\Man}^{(1:1)}=1$ and the corresponding effective saturated charge is
$\xi_{\eff}^{\text{sat}}=2/\pi\simeq 0.636$. 
We note that this result is in harmony with
exact bounds derived by Odijk in \cite{Odijk} :
$0.59 \xi_{\eff}^{\text{sat}}<0.67$.

This analysis clarifies the relationship between two different
notions, the Manning-Oosawa threshold for counterion condensation and
the notion of effective charge. The key point here is 
that the Manning-Oosawa
threshold $\xi_{\Man}$ for the bare charge is different from the value
of the saturated effective charge $\xi_{\eff}^{\text{sat}}$.
The fact that $\xi_{\eff}^{\text{sat}} < \xi_{\Man}$ is of course
a non-linear effect which means that even accounting correctly
for counterion condensation, the remaining layer of 
``uncondensed'' or ``free'' microions {\em cannot} be treated
within a linearized (Debye-H\"uckel-like) theory. This exact result
contrasts with common belief in the field, which possibly takes
its roots in the work of Manning \cite{Manning}.


\begin{figure}[htb]
\includegraphics[height=6cm,angle=0]{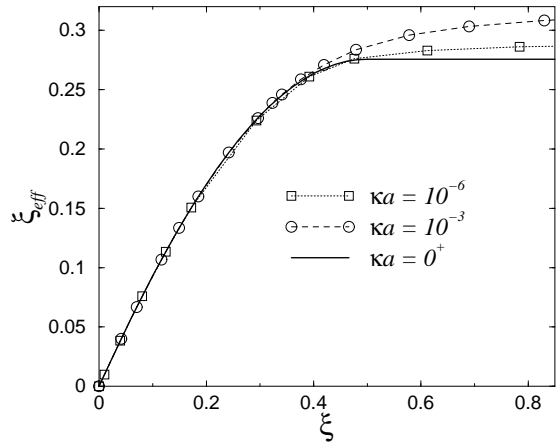}
\caption{Effective vs. bare line charge for an infinite charged 
rod (1:2 case).
The analytical result (\ref{eq:laeffkap012}) valid in the low salt or
thin rod limit $\kappa a \to 0$, shown by the thick continuous curve is 
compared to the numerical solution of Poisson-Boltzmann 
theory for $\kappa a=10^{-6}$ and 
$\kappa a=10^{-3}$. 
\Label{fig:laeffkap0}
}
\end{figure}


\begin{figure}[htb]
\includegraphics[height=6cm,angle=0]{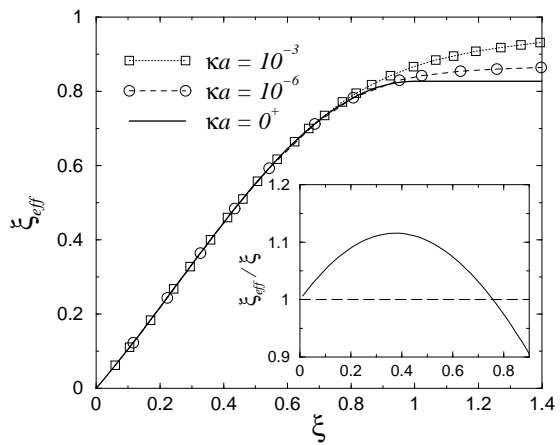}
\caption{
Same as Fig.~\ref{fig:laeffkap0} for the 2:1 electrolyte. Inset:
$\xi_{\eff}/\xi$ for small values of $\xi$. Notice the initial
overshooting effect $\xi_{\eff}>\xi$.
\Label{fig:laeffkap0-2:1}}
\end{figure}


In Figs.~\ref{fig:laeffkap0} and~\ref{fig:laeffkap0-2:1}, we compare
the result of Eqs.~(\ref{eq:laeffkap012}) and~(\ref{eq:laeffkap021})
for a 1:2 and 2:1 electrolyte, respectively, with numerical data, for
two low values of $\kappa a$. All numerical PB data have been 
obtained following the method discussed in \cite{HHH}.
One may conclude that the limiting
behavior $\kappa a \to 0$ is reached very slowly, in fact,
logarithmically, as we will show below.

It is interesting to notice in Fig.~\ref{fig:laeffkap0-2:1}, that in
the 2:1 electrolyte we have the overshooting effect, previously
reported in Ref.~\cite{Tellez-Trizac-PB-large-a}, where the effective
charge becomes larger than the bare charge for intermediate values of
the latter. Indeed for $|\xi| \ll 1$, from
Eqs.~(\ref{eq:laeffkap0-all}), we have for a 2:1 electrolyte
\begin{equation}
  \xi_\text{eff} = 
  \xi +
  \frac{\pi \xi^2}{3\sqrt{3}} 
  -\frac{2\pi^2 \xi^3}{27}  
  +{\cal O}(\xi^4)
  \qquad(2:1)
\end{equation}
and for a 1:2 electrolyte
\begin{equation}
  \xi_\text{eff} =
  \xi
  -\frac{\pi \xi^2}{3\sqrt{3}} 
  -\frac{2\pi^2 \xi^3}{27}  
  +{\cal O}(\xi^4)
  \qquad(1:2)
\end{equation}
The first deviation of the effective charge from linear behavior is
positive in the 2:1 case (overshooting effect) and negative in the 1:2
case (no overshooting). 
For a 1:1 electrolyte, 
\begin{equation}
  \xi_{\text{eff}} = 
  \xi -\frac{\pi^2}{24}
  \xi^3 
  +{\cal O}(\xi^5)
\,.
\end{equation}
The first deviation of the effective charge from the bare one is of
order $\xi^3$ (as in the case $\kappa a\gg
1$~\cite{Tellez-Trizac-PB-large-a}) and negative (thus no
overshooting). There is no term in $\xi^2$ as opposed to
the charge asymmetric electrolytes 1:2 and 2:1.
The physical origin of the overshooting effect
lies in the fact that in the 2:1 case, the divalent coions are expelled
further away in the double layer than the monovalent ones in the 1:1 situation, which
results in a stronger electrostatic potential at large distances. 
Similarly, the screening for 1:2 electrolytes is more efficient
since divalent counterions are available, which not only results
in the fact that $\xi_{\text{eff}}<\xi$, but also in the
lower value of $\xi_{\text{eff}}^{\sat}$. We therefore speculate that
the overshooting effect is generic in a $z$:$z'$ electrolyte, with 
coions of valency $z$ larger than that of counterions ($z'$).

\subsection{$\xi\to 0$ and arbitrary $\ha<1$}
In the limit where $\xi\to 0$, the solution (\ref{eq:DH-solut})
becomes exact for all distances, and one has $\xi_\eff/\xi \to 1$, as
was simply checked for $a = 0$ in section \ref{ssec:ha0}. At
finite salt however, the requirement $\xi_\eff/\xi \to 1$ when $\xi
\to 0$ provides quite a non trivial benchmark for our analytical
expressions.  It turns that the results of sections \ref{ssec:connect}
and \ref{ssec:bc} are not sufficient --at finite $\ha$-- to show
$\xi_\eff/\xi \to 1$. This is because in
Eqs.~(\ref{eq:small-asymptotics}) we neglected terms of order
$\hr^{2+2A}$ in the 1:1 and 1:2 cases and terms of order $\hr^{2+4A}$
in the 2:1 case. As $\xi\to 0$, we have $A\to0$, thus these terms
become of order $r^2$ and are of the same order of the last term in
the logarithms of Eqs.~(\ref{eq:small-asymptotics}). Then, to properly
compute the short-distance asymptotics as $\xi\to0$, we need the next
order terms in Eqs.~(\ref{eq:small-asymptotics}). These can be easily
obtained by replacing the small-$r$
asymptotics~(\ref{eq:small-asymptotics}) into the Poisson--Boltzmann
differential equation~(\ref{eq:PB}). For example for the 1:1 case, we
obtain, if $A<1$,
\begin{eqnarray}
    e^{-y_{11}/2}&=&\frac{\hr^A }{B}\Bigg[1-
    \frac{\hr^{2}}{16}\left(\frac{B^2\hr^{-2A}}{(1-A)^2}-
    \frac{\hr^{2A}}{B^2(1+A)^2}\right)
    \nonumber\\
    &&
    +{\cal O}(\hr^{2(2+2A)})
   \Bigg]
  \end{eqnarray}
Notice the expected symmetry in the two terms when one changes
$A\to-A$ (in which, from the definition of $B$, gives $B\to
B^{-1}$). Taking the limit $A\to0$ we obtain
\begin{equation}
  \label{eq:linear-limit}
  y_{11}(r)=-2A\left[\ln\frac{\hr}{2}+\gamma +
    \frac{\hr^2}{4}\left(\gamma+\ln\frac{\hr}{2}-1 \right)
    \right]+{\cal O}(\hr^4)
\end{equation}
We recognize the small-$r$ expansion of the Bessel function
$K_0(\hr)$.  On the other hand~(\ref{eq:connection11a}) says that for
$A\to0$, $A=2\lambda=\xi_{\eff}/(\ha K_1(\ha))$. We conclude that the
small-$r$ expansion~(\ref{eq:linear-limit}) is the small-$r$ expansion
of the linear solution~(\ref{eq:DH-solut}) and that, clearly, imposing
the boundary condition~(\ref{eq:BC}) at $r=a$ will yield
$\xi=\xi_{\eff}$.

A similar conclusion is reached for the asymmetric cases. For the 2:1
case the small-$r$ expansion, for $A<1$, reads
\begin{eqnarray}
  e^{-y_{21}/2}&=&\hr^{A}B^{-1/2}\Bigg[1
  -\frac{\hr^{2-2A}B}{24(1-A)^2} + \frac{\hr^{2+4A}}{24B^2(2A+1)^2}
  \nonumber\\
  &&+{\cal O}(\hr^{2(2+4A)})
  \Bigg]
  \label{eq:y21-small-A}
\end{eqnarray}
while the one for $y_{12}$ can be obtained from the preceding
expression through the substitution $A\to-A$ and $B\to B^{-1}$. It is
direct to verify that in the limit $A\to0$, the
asymptotics~(\ref{eq:y21-small-A}) yield
again~(\ref{eq:linear-limit}), that is the small-$r$ asymptotics of
$y_{12}$ and $y_{21}$ are those of the Bessel function $K_0(\hr)$,
solution of the linear problem.

\subsection{$\xi=\xi_\Man$ and arbitrary $\ha<1$}
\Label{ssec:ximan}
It is in general not possible to find analytically the solution 
to the transcendental equations satisfied by the parameter $\mu$,
which quantifies the behavior at short distances above
the salt dependent threshold, ie for $\xi>\xi_c$.
However, if $\xi=\xi_\Man$ (which is always beyond $\xi_c$), 
the relevant roots of Eqs. 
(\ref{eq:bc11b}), (\ref{eq:bc12b}) and (\ref{eq:bc21b})
read
\begin{subequations}
\Label{eq:muexact}
\begin{eqnarray}
\mu^{(1,1)} \,=\,-\frac{\pi}{4 [\log \ha + {\cal C}^{(1:1)}]}
\\
\mu^{(1,2)} \,=\,-\frac{\pi}{6 [\log \ha + {\cal C}^{(1:2)}]}
\\
\mu^{(2,1)} \,=\,-\frac{\pi}{3 [\log \ha + {\cal C}^{(2:1)}]}.
\end{eqnarray}
\end{subequations}
These expressions will be useful in section \ref{sec:discussion}.

\subsection{$\xi>\xi_\Man$ and arbitrary $\ha<1$}

The limit of large enough $\xi$ allows to derive a useful
approximation for $\mu$, that will turn important from a practical
point of view. Coming back to Eqs.~(\ref{eq:bc11b}), (\ref{eq:bc12b})
and (\ref{eq:bc21b}), we note that for diverging $\xi$, the arguments
of the tangent functions should be close to $-\pi$, so that the
tangent functions vanish, to be compatible with a finite value of
$\mu$. Expanding then the tangent to first order, we obtain
\begin{subequations}
\Label{eq:muapprox}
\begin{eqnarray}
\mu^{(1,1)} &\simeq&-\frac{\pi}{2 [\log \ha + {\cal C}^{(1:1)} -1/(\xi-1)]}
\label{eq:mu11approx}
\\
\mu^{(1,2)} &\simeq&-\frac{\pi}{3 [\log \ha + {\cal C}^{(1:2)}-1/(2\xi-1)]}
\\
\mu^{(2,1)} &\simeq&-\frac{2\,\pi}{3 [\log \ha + {\cal C}^{(2:1)}-1/(\xi-1)]}
\end{eqnarray}
\end{subequations}
The domain of validity of these relations is estimated to
be $\xi>\xi_\Man + {\cal O}(1/\log \ha)$, as may be observed
in Fig. \ref{fig:mu11} in the 1:1 case. In the vicinity 
of $\xi_\Man$ where (\ref{eq:muapprox}) fails, one should resort
to expressions (\ref{eq:muexact}).
We finally note that when $\xi$ diverges, $\mu$ saturates
to a finite value [twice those reported in (\ref{eq:muexact})]
\begin{subequations}
\Label{eq:musat}
\begin{eqnarray}
\Label{eq:musat11}
\mu^{(1,1)}_\sat \,=\,-\frac{\pi}{2 [\log \ha + {\cal C}^{(1:1)}]}
\\
\mu^{(1,2)}_\sat \,=\,-\frac{\pi}{3 [\log \ha + {\cal C}^{(1:2)}]}
\\
\mu^{(2,1)}_\sat \,=\,-\frac{2\,\pi}{3 [\log \ha + {\cal C}^{(2:1)}]}.
\end{eqnarray}
\end{subequations}
From Eq. (\ref{eq:connection11b}) and (\ref{eq:mulambda}) we conclude
that $\lambda$ does also saturate:
a corollary is that effective charges --directly related to $\lambda$
through (\ref{eq:xieff-lambda})-- saturate,
a generic feature of  mean-field theories 
\cite{PREsat}.

\section{Results at finite salt and discussion}
\Label{sec:discussion}

The results reported in sections \ref{sec:asymptotic} and
\ref{sec:limiting} are based on asymptotic expansions in 
the limit $\ha \to 0$. This limit is approached logarithmically
slowly, so that finite $\ha$ corrections should always be important
in practice. We now therefore address the question of the reliability of
our expressions at finite salt,
by confronting them with the numerical solution of the non-linear
PB equation (\ref{eq:PB}).

\subsection{Counterion condensation and Manning radius}

One of the most interesting features emerging here is that it is
possible to generalize the notion of counterion condensation
to finite salt systems. It is important however to emphasize here that 
condensation is not an all or nothing process,
that would occur precisely at $\xi_c$. Instead, it is 
a gradually built phenomenon. This is particularly true at 
finite salt density but already
in the limit of vanishing salt $\kappa a\to 0$,
it is noteworthy that non linear effects play an important role
for $\xi<\xi_c$, as revealed by the fact that $\xi_\eff/\xi \neq 1$.

We concentrate here on a 1:1 salt,
but the same analysis also holds for 1:2 and 2:1 salts.
The value $\xi_c$ discriminates between the short
distance behaviors (\ref{eq:small-asymptotics11}) for $\xi<\xi_c$
and (\ref{eq:pttedistgrdlambda11}) for $\xi>\xi_c$.
As may be observed in Fig. \ref{fig:A11}, $A$ is close to $\xi$ 
except in the vicinity of the threshold charge $\xi_c$, so that to
dominant order, $y_{11}$ behaves as $-2 \xi \log \hr$, which corresponds
to the potential of a bare cylinder with reduced charge $\xi$,
unaffected by screening. At $\xi=\xi_c$, the potential develops
an additional $\log\log r$ term (see section \ref{ssec:crit})
while for $\xi>\xi_c$, the behavior stemming from 
(\ref{eq:pttedistgrdlambda11}) is more complex. 
For high enough $\xi$ (more precisely, for 
$\xi>\xi_\Man + |{\cal O}(1/\log \ha)|>\xi_\Man$ (which is
itself larger than $\xi_c$), one may resort to approximation
(\ref{eq:muapprox}) and after some manipulations, rewrite 
(\ref{eq:pttedistgrdlambda11}) as 
\begin{equation}
e^{-y_{11}/2} \,\stackrel{r\simeq a}{\simeq} \, \frac{\hr}{2} \left[\log\left(
\frac{\hr}{\ha} \right)+\frac{1}{\xi-1}
\right].
\Label{eq:Rama}
\end{equation}
Here, we have used the fact that 
\begin{equation}
2 \mu \left(\log \hr + {\cal C}^{(1:1)}\right) = -\pi + 2 \mu \log\left(\frac{r}{a}\right)
+\frac{2 \mu}{\xi-1}
\end{equation}
so that Eq.~(\ref{eq:Rama}) holds for $\log(r/a) \ll {\cal O}(1/\mu)$.
Eq.~(\ref{eq:Rama}) had already been derived by Ramanathan
\cite{Ramanathan} (see also appendix A of \cite{NetzJ})
and implies that to dominant order, the potential behaves like 
$-2 \log r$ at short distances, which corresponds to the
bare potential of a polyion with $\xi=\xi_\Man=1$,
and is the fingerprint of counterion condensation.
We also note that for high $\xi$, Eq. (\ref{eq:Rama}) 
yields a total concentration of ions close to the rod that is
proportional to the square of the surface charge density,
as in the planar case \cite{Gueron}.
To complement the above results that assume
$\xi>\xi_\Man + |{\cal O}(1/\log \ha)|$, we also mention that for $\xi=\xi_\Man$
(ie above the threshold), the results of section 
\ref{ssec:ximan} yield
\begin{equation}
e^{-y_{11}/2} \,\stackrel{r\simeq a}{\simeq} \, -\frac{\hr}{\pi} 
\left(\log \ha +{\cal C}^{(1:1)} \right).
\end{equation}
The dominant behavior for $y_{11}$ is therefore again 
$-2 \log r$. Finally, decreasing further $\xi$ to investigate the
regime where it is close to the threshold $\xi_c$ (but
still larger than $\xi_c$), one may take advantage that $\mu$ vanishes
at $\xi=\xi_c$ to get Eqs.~(\ref{eq:criticalpot})
\begin{subequations}
\Label{eq:potcrit}
\begin{eqnarray}
e^{-y_{11}/2}&=& -\frac{\hr}{2} \,
\left(\log \hr +{\cal C}^{(1:1)} \right)
\\
e^{-y_{12}}&=& -\frac{\hr}{\sqrt{3}} \,
\left(\log \hr +{\cal C}^{(1:2)} \right)
\\
e^{-y_{21}/2}&=& -\frac{\hr}{\sqrt{6}} \,
\left(\log \hr +{\cal C}^{(2:1)} \right)
.
\end{eqnarray}
\end{subequations}
The dominant behavior at short scales is again $y \simeq -2 \xi_\Man
\log \hr$, which is the unscreened potential created by a rod of
reduced charge $\xi_\Man$.  This is quite remarkable given
$\xi_c<\xi_\Man$ since it holds in particular for a cylinder with
$\xi$ verifying $(\xi_c<)\xi<\xi_\Man$. Such a remark is nevertheless
quite misleading since at finite values of $\kappa a$, the requirement
$r>a$ does not allow to take the limit $r \to 0$ where the different
leading and sub-leading contributions to the potential can be
identified: ``corrections'' to the ``leading'' term are important.
In other words, the full expression is required to approximate $y$
in (\ref{eq:potcrit}) and the ``dominant'' term 
$-2 \xi_\Man\log \hr$ is in practice a bad approximation.

Our analytical expressions identify a mathematical change in the
behavior of the electric potential at $\xi=\xi_c$, that may be
considered as being associated with
counterion condensation. Such a terminology may however be confusing
since at finite $\ha$, no singularity signals the crossing of
the threshold value $\xi_c$. This restriction should be borne in mind. 
Of particular interest is the condensate
structure, completely encoded in Eqs.~(\ref{eq:pttedistgrdlambda}),
and in particular the condensate thickness \cite{Deserno,OS}. Its
definition is necessarily arbitrary.  The so-called Manning radius is
often considered when it comes to quantifying the condensate size. It
is defined as the distance $r_m$ where the integrated charge $q(r) =
-(r/2) dy/dr$ equals $\xi_\Man$. Since we have seen that
$\xi_c<\xi_\Man$, it is clear that such a criterion in inoperant for
$\xi_c<\xi<\xi_\Man$, which constitutes quite a deficiency, but the
corresponding value $r_m$ nevertheless exhibits remarkable features
within its domain of definition $\xi>\xi_\Man$.  We first compute
$q(r)$, that follows directly from Eqs. (\ref{eq:pttedistgrdlambda})
\begin{subequations}
\Label{eq:qder}
\begin{eqnarray}
q^{(1:1)}(r) &=& 1 + 2\mu \,\tan\left[
-2 \mu \log\left(\frac{r}{r_m^{(1:1)}}\right)
\right]
\\
q^{(1:2)}(r) &=& \frac{1}{2} + \frac{3\mu}{2} \,\tan\left[
-3 \mu \log\left(\frac{r}{r_m^{(1:2)}}\right)
\right]
\\
q^{(2:1)}(r) &=& 1 + \frac{3\mu}{2} \,\tan\left[
-\frac{3 \mu}{2} \log\left(\frac{r}{r_m^{(2:1)}}\right)
\right].
\end{eqnarray}
\end{subequations}
with
\begin{subequations}
\Label{eq:OSmieux}
\begin{eqnarray}
\kappa r_m^{(1:1)} &=& \exp\left(-{\cal C}^{(1:1)} -\frac{\pi}{4 \mu}
\right)
\Label{eq:OSmieux11}
\\
\kappa r_m^{(1:2)} &=& \exp\left(-{\cal C}^{(1:2)} -\frac{\pi}{6 \mu}
\right)
\Label{eq:OSmieux12}
\\
\kappa r_m^{(2:1)} &=& \exp\left(-{\cal C}^{(2:1)} -\frac{\pi}{3 \mu}
\right)
\Label{eq:OSmieux21}
\end{eqnarray}
\end{subequations}
In these expressions, one observes that $q(r)=\xi_\Man$ for
$r=r_m$, which means that the $r_m$ appearing in (\ref{eq:qder})
are by definition Manning
radii. When $\xi=\xi_\Man$, then by definition
the Manning radius should coincide with the polyion radius $a$.
This is indeed the case here, as a consequence of Eqs. (\ref{eq:muexact}).
For large enough $\xi$, we obtain an expression in closed
form from approximation (\ref{eq:muapprox}):
\begin{subequations}
\Label{eq:OS}
\begin{eqnarray}
\kappa r_m^{(1:1)} &\simeq & 2 \sqrt{2 \kappa a} \, \exp\left(
-\frac{\gamma}{2} -\frac{1}{2(\xi-1)}
\right)
\Label{eq:OS11}
\\
\kappa r_m^{(1:2)} &\simeq& 
\frac{3^{3/4}}{2^{1/3}} \sqrt{2\kappa a} \, \exp\left[
-\frac{\gamma}{2} -\frac{1}{2(2\xi-1)}
\right]
\\
\kappa r_m^{(2:1)} &\simeq& 3^{3/4} \sqrt{2\kappa a} \, \exp\left[
-\frac{\gamma}{2} -\frac{1}{2(\xi-1)}
\right].
\end{eqnarray}
\end{subequations}
In the cell model, the previous definition 
\begin{equation}
q(r_m) = \xi_\Man
\Label{eq:rm}
\end{equation}
offers a geometric construction to compute $r_m$: $q(r)$ plotted as a
function of $\log r$ displays an inflection point at $r=r_m$ (see Ref.
\cite{Deserno} where it was also shown that this criterion could be
extended beyond mean-field and used in Molecular Dynamics simulations to
define a fraction of condensed ions). From the functional form of
Eqs.~(\ref{eq:qder}) --with a shifted tangent as a function of $\log
r$, i.e. the same form as in the salt-free cell model case-- we see
that a similar inflection point criterion holds here.  This will be
illustrated further in section \ref{ssec:pot}.

For a 1:1 electrolyte, (\ref{eq:OS11}) is precisely the result obtained
in Ref. \cite{OS}. The present work therefore precises the domain of validity
of this expression, and offers with (\ref{eq:OSmieux11}) supplemented 
with (\ref{eq:bc11b}) a better approximation
(see Figure \ref{fig:rm11}). By construction, the accuracy of
our expressions improve upon decreasing $\ha$. To test the worst cases,
we therefore chose
a relatively ``high'' value $\ha = 0.01$ in Fig.~\ref{fig:rm11},
where the agreement is seen to be very good, while (\ref{eq:OS11})
fails when $\xi$ is too close to $\xi_\Man=1$. Fig.~\ref{fig:rm21}
illustrates the effect of increasing salinity and displays the
results associated to three values of $\ha$. It is observed 
that at high $\xi$, the analytical predictions deviate all the more
from the numerical results computed for PB theory as $\ha$ is
increased. For the highest value $\ha=0.3$ of Fig. \ref{fig:rm21},
the error made when $\xi \to \infty$ is on the order of 30\%.
It also appears that even at relatively high $\ha$, the analytical 
prediction is reliable for low $\xi$ [bearing in mind
that $\xi$ must be larger than $\xi_\Man$ to allow for the
definition of $r_m$ through Eq. (\ref{eq:rm})].
For completeness, we present results for the 1:2 case as well
in Figure \ref{fig:rm12}, changing $\ha$ at fixed charge
$\xi=0.55$ which is just 
above the value $\xi_\Man=1/2$. The results at saturation 
are also reported (dashed line and squares).
One needs to push $\ha$ beyond 0.1 to notice
a difference between (\ref{eq:OSmieux12}) and the numerical
PB result (see the inset). Figure \ref{fig:rm12}
illustrates how finite salt effects influence the scaling
exponent $\alpha$ in the relation $\kappa r_m \propto (\kappa a)^\alpha$.
If $\xi > \xi_\Man + |{\cal O}(1/\log \ha)|$,
that is at high enough $\xi$ or low enough $\ha$, 
we have $\alpha=1/2$ as dictated by Eqs. (\ref{eq:OS})
and evidenced by the dotted line in the main graph.
If on the other hand one sits in  the region 
$\xi_\Man< \xi < \xi_\Man + |{\cal O}(1/\log \ha)|$,
the exponent $\alpha$ changes continuously, 
and increasing $\ha$ at fixed $\xi$ drives the system toward the
regime $\alpha=1$, a feature --visible for the data at $\xi=0.55$ in Fig.
\ref{fig:rm12}-- which simply reflects the fact that $r_m \to a$.
Finally, we emphasize that the behavior in 
all three 1:1, 1:2 and 2:1 situations are qualitatively very similar
so that the conclusions reached and phenomena observed
are transferable from one situation to another.

\begin{figure}[htb]
\includegraphics[height=6cm,angle=0]{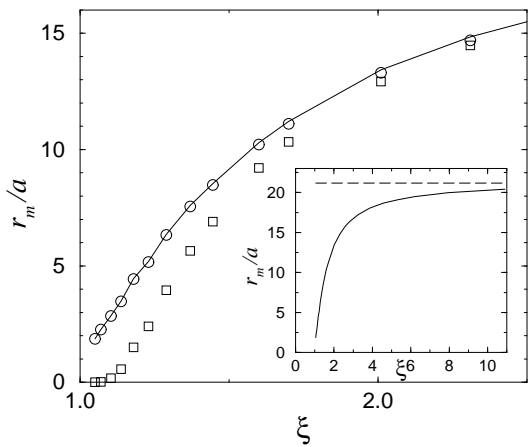}
\caption{Manning radius $r_m$ as defined from Eq. (\ref{eq:rm}) 
versus reduced bare charge $\xi$, for a 1:1 salt with $\kappa a=10^{-2}$
($a$ is the polyion radius).
The continuous curve shows the results obtained from the numerical
solution of PB equation, while the circles stand for expression
(\ref{eq:OSmieux11}) where $\mu$ is the smallest positive root of
(\ref{eq:bc11b}) (see Fig. \ref{fig:mu11} for a plot of $\mu$ as a function of
bare charge).
The squares indicate the prediction (\ref{eq:OS11})
(see the dashed line in Fig. \ref{fig:mu11} for the associated
$\mu$)
which becomes asymptotically correct for large $\xi$. The inset
displays the corresponding charge regime, where $r_m/a$ saturates 
at high $\xi$ to
the value $2\sqrt{2/\ha} \exp(-\gamma/2)$ indicated by the dashed horizontal line.
\Label{fig:rm11}
}
\end{figure}

\begin{figure}[htb]
\includegraphics[height=6cm,angle=0]{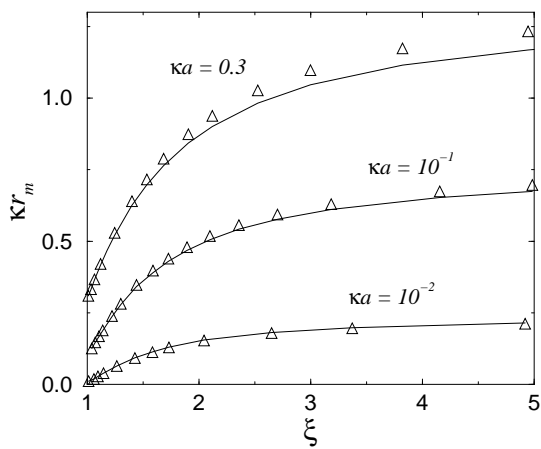}
\caption{The Manning radius $r_m$ versus the reduced bare charge $\xi$
for a 2:1 salt, for different values of $\kappa a$. The continuous
curve shows the prediction of Eq.~(\ref{eq:OSmieux21}) while the
triangles correspond to the numerical solution of PB theory.
\Label{fig:rm21}
}
\end{figure}

\begin{figure}[htb]
\includegraphics[height=6cm,angle=0]{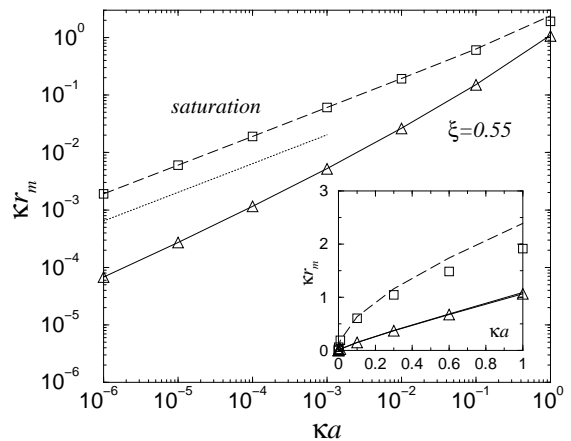}
\caption{Manning radius as a function of salinity $\ha$ for two
different charges and a 1:2 salt. 
The symbols show the prediction of Eq. (\ref{eq:OSmieux12})
(triangles for $\xi=0.55$ and squares in the saturation limit
corresponding to $\xi\to \infty$). The continuous and dashed lines 
display the corresponding 
numerical solution of PB theory. The dotted line is a guide for
the eye indicating a slope $\alpha=1/2$.
The inset shows the same
data on a linear scale.
\Label{fig:rm12}
}
\end{figure}

The inflection point feature stemming from (\ref{eq:rm})
has the merit to unify the present 
infinite dilution/finite salt phenomenology with 
the finite density/vanishing salt
situation of the cell model. 
It should be kept in mind however that (\ref{eq:rm}) does not allow
to define a Manning radius for $\xi_c<\xi<\xi_\Man$, where 
the condensation phenomenon is already present, {\em at least from
a mathematical point of view} with a change in the short 
distance behavior of $y(r)$. In addition the scaling $r_m \propto
(a/\kappa)^{1/2}$ valid for large enough $\xi$ [see (\ref{eq:OS})]
is definition dependent. An alternative to (\ref{eq:rm})
could be to define a characteristic radius of the condensate 
through $q(r^*) = \xi_c$. Making use of (\ref{eq:muapprox})
then implies that for a 1:1 salt
$\kappa r^* \propto (\kappa a)^\alpha$
with $\alpha = (\hbox{arctan}\, \pi)/\pi \simeq 0.402$ instead of $\alpha=1/2$
in (\ref{eq:OS}). As (\ref{eq:muapprox}), this is limited to high 
enough $\xi$. For $\xi=1$, the results of section \ref{ssec:ximan}
allow for an analytical computation of $r^*$ and yield
$\alpha = (2/\pi)\, \hbox{arctan}(\pi/2) \simeq 0.639$.

Our expressions
also allow to discuss several quantities that directly follow
from the electric potential, such as the Bjerrum radius $r_B$
considered in Ref. \cite{Qian} and defined as the locus of
an inflection point in the integrated counterion density 
when plotted as a function of radial distance. 
For 1:1 and 2:1 salts, this definition implies $q(r_B) = 1/2$
while in the 1:2 case, we have $q(r_B)=1/4$. Making use of 
Eqs. (\ref{eq:qder}) provide a transcendental equation 
from which $r_B$ follows.

\subsection{Effective charges}
\Label{ssec:xisat}

Of particular interest to describe interactions at large distances
(typically $r > \kappa^{-1}$ as will be discussed in section 
\ref{ssec:pot}) is the effective charge defined from 
the far field asymptotics (\ref{eq:DH-solut}). This quantity
is given by Eqs. (\ref{eq:xieff-lambda}) where $\lambda$ follows
from the expressions given in sections \ref{ssec:connect} and 
\ref{ssec:bc}. The results pertaining to the limit $\ha \to 0$ have
been given in section \ref{ssec:ha0}, but it is also possible
to derive closed form relations in the saturation limit $\xi \to \infty$
where $\mu$ is given by (\ref{eq:musat}). We therefore have
(denoting $\xi_\eff^\sat$ by $\xi_\sat$):
\begin{widetext}
\begin{subequations}
\Label{eq:xisat}
\begin{eqnarray}
\xi^{(1:1)}_\sat &=&\ha K_1(\ha)\,\frac{2}{\pi} \,\left[
\cosh\left(\frac{\pi^2}{2 [\log \ha + {\cal C}^{(1:1)}]}
\right)\right]
\Label{eq:xisat11}
\\
\xi^{(1:2)}_\sat &=&\ha K_1(\ha)\,\frac{\sqrt{3}}{\pi} \,\left[
\cosh\left(\frac{\pi^2}{3 [\log \ha + {\cal C}^{(1:2)}]}
\right)-\frac{1}{2}\right]
\Label{eq:xisat12}
\\
\xi^{(2:1)}_\sat &=&\ha K_1(\ha)\,\frac{\sqrt{3}}{\pi} \,\left[
\cosh\left(\frac{2\,\pi^2}{3 [\log \ha + {\cal C}^{(2:1)}]}
\right)+\frac{1}{2}\right]
\Label{eq:xisat21}
\end{eqnarray}
\end{subequations}
\end{widetext}

\begin{figure}[htb]
\includegraphics[height=6cm,angle=0]{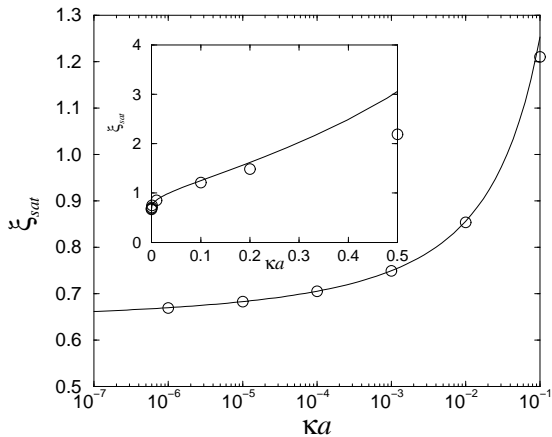}
\caption{
Effective linear charge density at saturation,
$\xi_{\text{sat}}$ as a function of the radius $\ha$ of the
rod for a symmetric 1:1 electrolyte. The full line is the
analytical expression~(\ref{eq:xisat11}), the circles have been
obtained by solving numerically Poisson--Boltzmann equation.
The inset shows the same data on a linear scale.
\Label{fig:lasat11}
}
\end{figure}

Figures \ref{fig:lasat11} and \ref{fig:lasat} show that the analytical
expressions (\ref{eq:xisat}) are in good agreement with the
numerical results for $\kappa a=\ha<0.1$. For $\ha>0.1$, deviations
become apparent (see the inset of Fig. \ref{fig:lasat11}).
We will propose in section \ref{sec:upperbound} an alternative approach that 
covers the whole range of $\kappa a$.

\begin{figure}[htb]
\includegraphics[height=6cm,angle=0]{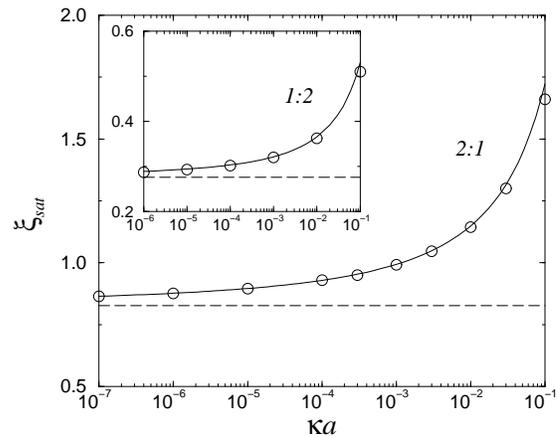}
\caption{
Same as the main graph of Fig. \ref{fig:lasat11} for 1:2 and 2:1
electrolytes. The analytical expressions shown by the lines
are those of Eqs. (\ref{eq:xisat12}) and (\ref{eq:xisat21}).
The limiting values for $\kappa a \to 0^+$ are shown by the
dashed lines.  
\Label{fig:lasat}
}
\end{figure}

We have also checked that the effective charges are correctly 
described by the analytical predictions of section \ref{sec:asymptotic},
not only in the saturation regime but for arbitrary values
of the charge $\xi$ provided $\ha <0.1$.

\subsection{Electric potential}
\Label{ssec:pot}

\begin{figure}[htb]
\includegraphics[height=6cm,angle=0]{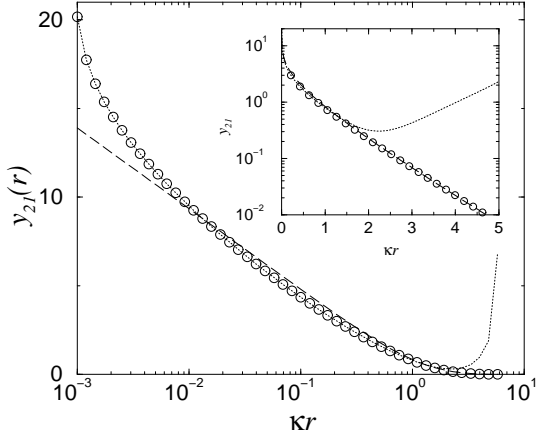}
\caption{Electrostatic potential in a 2:1 electrolyte as a function of
radial distance for a rod with a high bare charge ($\xi \simeq 11$)
and $\kappa a=10^{-3}$. The circles represent the numerical solution of
PB theory, the dotted line is for the short distance formula
(\ref{eq:pttedistgrdlambda21}) and the dashed line is for the 
far field expression (\ref{eq:DH-solut}). In the inset, the same
results are shown on a linear-log scale.
\Label{fig:pot211e-3}
}
\end{figure}

\begin{figure}[htb]
\includegraphics[height=6cm,angle=0]{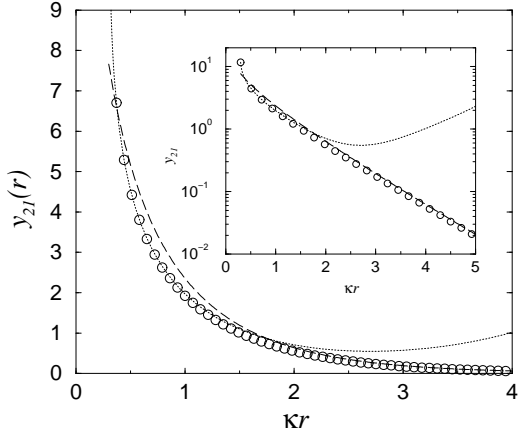}
\caption{Same as Fig. \ref{fig:pot211e-3} with $\xi \simeq 40$
and $\kappa a=0.3$, on a linear scale. 
\Label{fig:pot210.3}
}
\end{figure}

So far, we focused on several scalar quantities characterizing
the electric potential, and thus the ionic distribution. It is
also instructive to compare the analytical potentials to their
``exact'' PB counterparts. When $\xi>\xi_c$, the relevant short scale
expressions are those of Eqs. (\ref{eq:pttedistgrdlambda}).
We take the 2:1 situation as an illustrative example.
As expected from the results of section \ref{ssec:xisat},
Figures \ref{fig:pot211e-3} and \ref{fig:pot210.3} show that 
the far field behavior is well captured by (\ref{eq:DH-solut}),
and it also appears that such an expression may be used 
in practice for $\kappa r>1$. On the other hand, expansion 
(\ref{eq:pttedistgrdlambda21}) is very accurate at small distances,
and may be extended up to $\kappa r$ of order 1. For  the low values
of $\kappa a$ such as that of Fig. \ref{fig:pot211e-3}, expansion
(\ref{eq:pttedistgrdlambda21}) produces the correct potential 
with an exceptional accuracy for several orders of magnitude in $r$.
For higher values of $\kappa a$, the short distance expansion
(\ref{eq:pttedistgrdlambda21}) is still useful and reliable
up to $\kappa r$ of order unity (see Fig. \ref{fig:pot210.3}).
To illustrate the improvement over previous expansions,
we come back to the 1:1 case in Fig. \ref{fig:pot11} 
and compare our formula
with the classic one of Ramanathan \cite{Ramanathan}, recalled 
in (\ref{eq:Rama}) and re-derived in the present work.
Once the potential is known, one can compute the integrated
charge $q(r) = -(1/2) r dy/dr $, displayed in Fig. 
\ref{fig:qder11}. The coincidence between the $q$ versus
$\log r$ inflection point and the Manning radius 
may be observed in the inset. The integrated charge 
--shown by the dotted line-- following
from (\ref{eq:Rama}) cannot reproduce such a feature, 
and we also observe that the most significant decay
of $q(r)$ from its initial value ($\xi$ as requested by Gauss theorem,
i.e. 25 in Figure \ref{fig:qder11}) to unity occurs on a 
much shorter scale than the Manning radius.

\begin{figure}[htb]
\includegraphics[height=6cm,angle=0]{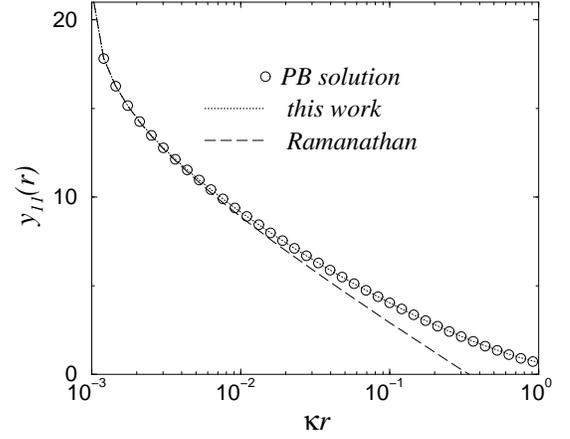}
\caption{Comparison of the numerical solution of PB theory
with expansion (\ref{eq:pttedistgrdlambda11}) shown by the dotted line and
with the formula of Ramanathan [see (\ref{eq:Rama}) and dashed line]
for $\kappa a =10^{-3}$, $\xi= 25$ and a 1:1 salt.
\Label{fig:pot11}
}
\end{figure}

\begin{figure}[htb]
\includegraphics[height=6cm,angle=0]{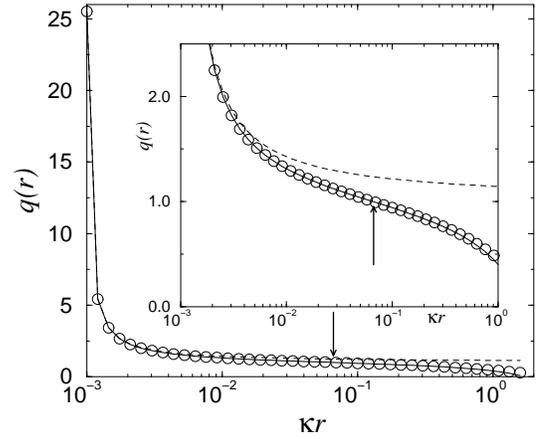}
\caption{Integrated charge as a function of rescaled distance for the same
parameters as in Fig. \ref{fig:pot11} (symbols and lines have the same
meaning here). The inset shows the same data in the vicinity of the predicted
Manning radius, indicated by the arrow. The prediction follows 
from (\ref{eq:OSmieux11}) or equivalently 
(\ref{eq:OS11}) since we consider here a highly charged rod.
\Label{fig:qder11}
}
\end{figure}

\section{Effective charge at saturation and the largest eigenvalue of
the operators $K_{\ha}$}
\Label{sec:upperbound}

The results of the previous sections are broadly speaking restricted
to the regime $\kappa a<1$. For the specific
problem of the effective charge in the 1:1 situation, 
it is however possible to obtain expressions that 
hold for all salinities.

The effective charge at saturation, for any arbitrary value of $\ha$,
has an interesting relation with the largest eigenvalue of the
operators defined in Eqs.~(\ref{eq:opK11}) and~(\ref{eq:opK12-21}). To
see this, consider first the 1:1 electrolyte, and the
solution~(\ref{eq:sol11}) to Poisson--Boltzmann equation. As the bare
charge $\xi$ increases the parameter $\lambda$ (related to the
effective charge $\xi_{\eff}$) increases. At saturation ($\xi\to
+\infty$), $\lambda=\lambda_{\sat}$ is such that the electric field
obtained from the solution~(\ref{eq:sol11}) diverges at
$\hr=\ha$. Clearly this happens if $\det(1-\lambda K_{\ha})=0$, thus
if $\lambda$ is the inverse of an eigenvalue of
$K_{\ha}$\footnote{This analysis is for (positive) saturation when
$\xi>0$. In the case $\xi<0$, the saturation is reached when
$\det(1+\lambda K_{\ha})=0$.}. Now, since when $\xi=0$, $\lambda=0$ and
it increases as $\xi$ increases, it appears that at saturation
$\lambda=\lambda_{\sat}$ is equal to the inverse of the largest
eigenvalue of $K_{\ha}$.

The same analysis applies to the 1:2 and 2:1 electrolytes. For a 1:2
electrolyte, for positive saturation $\xi\to+\infty$, $\lambda_{\sat}$
is the inverse of the largest eigenvalue of $K_{\ha}^{(2)}$, whereas
for a 2:1 electrolyte, when $\xi\to+\infty$, $\lambda_{\sat}$
is the inverse of the largest eigenvalue of $-K_{\ha}^{(0)}$.

We have not been able to find explicitly the eigenvalues of the
operators $K_{\ha}$, $K_{\ha}^{(0)}$ and $K_{\ha}^{(2)}$, for any
arbitrary value of $\ha$. However one can use approximate methods to
find estimates for the largest eigenvalue.

For the 1:1 electrolyte, it is shown 
in the appendix of~\cite{Tracy-asympot-tau},
that for any function $\phi\in
L^{2}(0,\infty,e^{-\ha(u+u^{-1})/2}\,du)$, the largest eigenvalue
$\lambda_{\sat} ^{-1}$ of $K_{\ha}$ satisfies
\begin{equation}
\frac{(\phi,K_{\ha}\phi)}{(\phi,\phi)}\leq \lambda_{\sat}^{-1}
\end{equation}
where $(\cdot\,,\cdot)$ is the scalar product of
$L^{2}(0,\infty,e^{-\ha(u+u^{-1})/2}\,du)$. Using the test function
$\phi(u)=1/\sqrt{u}$, we obtain
\begin{equation}
\lambda_{\sat}\leq \frac{K_0(\ha)}{\pi \Gamma(0,2\ha)}
\end{equation}
with $K_0$ the modified Bessel function and
$\Gamma(0,z)=\int_{z}^{+\infty}e^{-t}/t\,dt$. Using
Eq.~(\ref{eq:xieff-lambda11}), this finally gives an upper bound for
the effective charge at saturation, for a 1:1 electrolyte, for any
arbitrary value of $\ha$, $\xi_{\eff}^{\sat}\leq\xi^{\text{sat, up}}$,
with
\begin{equation}
\Label{eq:xi-upper-bound}
\xi^{\text{sat, up}}
=
\frac{2aK_{1}(\ha)K_0(\ha)}{\pi\,\Gamma(0,2\ha)}
\end{equation}
As $\ha\to 0$, $\xi^{\text{sat, up}} \to 2/\pi$, thus has the same
limit as the exact $\xi_{\eff}^{\sat}$. This is expected since in the
appendix~\cite{Tracy-asympot-tau}, this upper bound was used to prove
that the supremum of the largest eigenvalue of $K_{\ha}$ as $\ha\to 0$
is $\pi$.

Interestingly, when $\ha\gg 1$ we have
\begin{equation}
\xi^{\text{sat, up}}=2\ha+\frac{3}{2}+{\cal O}(\ha^{-1})
\end{equation}
which is the same asymptotic behavior of the true effective saturated
charge $\xi_{\eff}^{\sat}$ obtained in~\cite{ATB} by a different
approach than the present one. It is the effective charge of 
an infinite plane ($2 \kappa a$) plus a correction ($3/2$) that is obtained
using a small curvature expansion. 

\begin{figure}[t]
\includegraphics[height=6cm,angle=0]{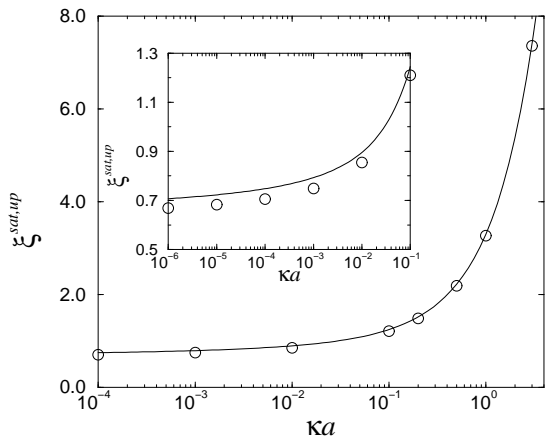}
\caption{ The upper bound~(\ref{eq:xi-upper-bound}) used as an ansatz
for the effective charge at saturation as a function of the radius
$\ha$ of the rod for a symmetric 1:1 electrolyte. The full line is
the analytical expression~(\ref{eq:xi-upper-bound}), while the circles have
been obtained by numerical resolution of Poisson--Boltzmann equation.
\label{fig:lasat_11_gde_val_propre}
}
\end{figure}

Figure~\ref{fig:lasat_11_gde_val_propre} shows a comparison between
$\xi^{\text{sat, up}}$ from Eq.~(\ref{eq:xi-upper-bound}) and the
effective charge at saturation $\xi_{\eff}^{\sat}$ obtained
numerically. Surprisingly, it turns out $\xi^{\text{sat, up}}$ is not
only an upper bound for $\xi_{\eff}^{\sat}$ but also a very good
estimate for $\xi_{\eff}^{\sat}$ for any value of $\ha$. If
$\ha>10^{-1}$ it is in very good agreement with the numerical
data. However, for very small radius $\ha\ll 1$, the
estimate~(\ref{eq:xisat11}) of section \ref{ssec:xisat}, is better than the
\textit{ansatz}~(\ref{eq:xi-upper-bound}) (see the inset of
Fig.~\ref{fig:lasat_11_gde_val_propre} and compare it to
Fig.~\ref{fig:lasat11}).


\section{Conclusion}
\Label{sec:concl}

The mathematical results derived in the framework of 
Painlev\'e/Toda type equations provides much insight into the
screening behavior of microions in the vicinity of a charged
rod-like macroion. Although the mapping between the Poisson-Boltzmann
equation for a certain class of electrolytes (1:1, 1:2 and 2:1) onto
a Painlev\'e/Toda equation is in itself not new, previous analysis
were concerned with the no salt limit where the ratio of
macroion radius $a$ to Debye length $\kappa^{-1}$ vanishes, and the practical 
consequences and implications of a finite salt concentration 
had not been drawned. We have shown here that systematic
logarithmic, and thus strong,
corrections arise for various quantities of interest.
All finite salt results reported here are new, 
and significantly improve previously available expressions.
In addition, the 2:1 situation worked out here had not been
considered before, even in the limit of vanishing salt content.

The experimental relevance of our findings has been sketched
in the introduction. A rough criterion for the validity of our mean-field
approach is that  $\Gamma=z^{3/2} \sqrt{\xi \ell_B/(2 \pi a)}<2$
where $z$ denotes the valency of counter-ions. For 1:1
and 2:1 electrolytes, the criterion is fulfilled even by the most 
highly charged polymers, such as double-stranded DNA
for which $a \simeq 1\,$nm and $\xi\simeq 4$, at least in
water where $\ell_B \simeq 0.7\,$nm. In the 1:2 case, 
we obtain for dsDNA $\Gamma \simeq 2$,
for which it is difficult to weight the importance of microionic
correlation against the mean-field effects reported here. 
In any case, our results apply to single stranded DNA
and polymers of lower line charge such as hyaluronan
\cite{Boue}.
An upper limit
should also be put to salt content ($\kappa\, \ell_B<1$),
but such a requirement is irrelevant for our purposes
since only the low salt regime has been discussed.

Section \ref{sec:asymptotic} contains the core
of our analysis.
It appeared there that the short scale behavior of the electric
potential depends on the position of the reduced bare charge $\xi$
with respect to a threshold $\xi_c$. One may consider that
this threshold is a remnant of its vanishing salt counterpart,
which signals the
onset of counter-ion condensation. It should be kept in mind
however that counter-ion condensation is not an all or nothing
process, but a gradually built phenomenon.

The charge $\xi_c$ is strongly salt dependent, and coincides with
Manning critical value in the limit $\kappa a \to 0$.
On the other hand, the large distance behavior is always of the
same functional form [see (\ref{eq:DH-solut})] and is governed by the
effective charge $\xi_{\eff}$. This latter quantity depends on a
parameter $\lambda$ that also plays a crucial role in the
description of the short distance behavior. We have proposed 
a simple approximation to compute explicitly $\lambda$
for 1:1, 2:1 and 1:2 electrolytes,
from which all other quantities follow.
The analytical predictions have been tested against 
the numerical solution of Poisson-Boltzmann theory,
and shown to be remarkably accurate
[of particular practical interest for highly charged polyelectrolytes
are Eqs.(\ref{eq:pttedistgrdlambda})  and (\ref{eq:DH-solut})
with $\mu$ given by (\ref{eq:muapprox}) and
where $\lambda$ is given as a function of $\mu$ 
in section \ref{ssec:connect}]. 
In particular, simple analytical
results have been derived for the Manning radius, that is often
used to quantify the lateral extension of the condensate that may
form around a cylindrical polyion. A few other measures of
the condensate thickness have been commented upon. 

Our approach is free of the matching procedures  \cite{Mac,Philip,vK}
or ad-hoc though educated assumptions \cite{Tuinier} underlying
previous work, and therefore provides expansions
with controlled error for the electric potential. 
The analysis of sections \ref{sec:asymptotic} to \ref{sec:discussion}
requires that $a<\kappa^{-1}$. For larger salinities, contributions
that have been neglected here become relevant. However, the spectral
analysis of section \ref{sec:upperbound} provides for all values of 
$\kappa a$ an upper bound
for the far field signature of the polyion, from which an 
excellent approximation of the effective charge at saturation may be
obtained.

The analytical results obtained emphasize the essential
difference between the critical Manning charge for condensation and 
effective charge of the macroion, the latter being defined
from the large scale electric behavior. A consequence is that 
within a simplified two state model where the population of 
microion is divided into a condensed region and a free population, 
free ions cannot be treated within 
a linearized theory, at variance with common belief and
practice. Consistency with exact results requires that
non linear effects are still at work in the free region
and significantly decrease the effective charge compared to 
the critical one. In the 1:2 case at ultra low salt, 
for all bare charges $\xi$ larger than $\xi_c$, the ratio
of $\xi_{\text{eff}}$ to $\xi_c$ (with $\xi_c$ equal to $\xi_\Man$ here
since $\kappa a \to 0$) is equal to $\sqrt{3}/\pi\simeq 0.55$.
This ratio is closer to unity for 2:1 electrolytes 
[$3 \sqrt{3}/(2\pi)\simeq 0.83$], and intermediate for a 1:1 salt
($2/\pi\simeq 0.64$). We therefore emphasize that 
the distinction between bound and free ions
is not only arbitrary but also potentially misleading.


\begin{acknowledgments}
We would like to thank T. Odijk for useful remarks.
This work was supported by a ECOS Nord/COLCIENCIAS action of French
and Colombian cooperation. G.~T.~acknowledge partial financial support
from COLCIENCIAS grant 1204-05-13625 and from Comit\'e de
Investigaciones de la Facultad de Ciencias de la Universidad de los
Andes. This work has been supported in part by the NSF PFC-sponsored Center              
for Theoretical Biological Physics (Grants No. PHY-0216576 and                    
PHY-0225630). 
\end{acknowledgments}

\begin{appendix}
\section{Painlev\'e classification, a brief reminder}
\Label{app:A}
Polynomial non-linear differential equations of the form 
\begin{equation}
A(x,y) \frac{d^2y}{dx^2} \,+\, B(x,y) \frac{dy}{dx}\,+\,
C(x,y) \left(\frac{dy}{dx}\right)^2 \,+\,D(x,y) \,=\, 0,
\end{equation}
where the functions $A,B,C,D$ are polynomial in $y$ and analytic in $x$,
have been classified with respect to the character of the singular points 
of the solutions. Of special interest are the equations for which 
branch points and essential singularities do not depend on initial 
conditions (hence the only movable singularities are poles).
Fifty canonical types of equations with the above property have
been uncovered, most of which (44) are integrable in terms of elementary 
functions. Solving the remaining 6 types requires the introduction of new
(Painlev\'e) transcendental functions. The third member 
(Painlev\'e III) of this family
of 6 corresponds to the generic form
\begin{equation}
x y  \frac{d^2y}{dx^2} \,=\, x \left(\frac{dy}{dx}\right)^2 
\,- \,y \frac{dy}{dx} \,+ \,a x +b y + c y^3 +d x y^4.
\end{equation}

\section{Short distance behavior for $\lambda>\lambda_c$}
\Label{app:B}

For a highly charged rod, when $\xi>\xi_c$, we have $\Re
e(A)=\xi_{\Man}$. Then, the last ``higher'' order term in
Eqs.~(\ref{eq:small-asymptotics}) become of the same order as the
${\cal O}(1)$ terms. Thus the small-$\hr$ asymptotics of $y(\hr)$ are
different, when $\xi>\xi_c$, than the ones given by
Eqs.~(\ref{eq:small-asymptotics}).

In~\cite{McCoy-Tracy-Wu} the asymptotics for $\lambda>1$ in the 1:1
case were studied and in~\cite{TracyWidom-Toda-asympt} the ones for
the 1:2 case for $\lambda>\lambda_c$ were obtained. We will not
reproduce those calculations here, but to illustrate the method
from~\cite{McCoy-Tracy-Wu,TracyWidom-Toda-asympt}, we will compute the
small-$\hr$ asymptotics for the 2:1 case, for
$\lambda>\lambda_c^{(2:1)}$, which has not been previously considered.

The asymptotic form~(\ref{eq:small-asymptotics21}) is actually valid even
if $\lambda$ is complex, provided that it
satisfies~(\ref{eq:lambda-conditions}). To study the asymptotics when
$\lambda>\lambda_c$ we can consider that $\lambda$ approaches the cut
$[\lambda_c,+\infty)$ from below, for instance. Then we can
rewrite~(\ref{eq:A12}) as
\begin{equation}
A=1-\frac{3}{2}i\mu
\end{equation}
with
\begin{subequations}
\begin{eqnarray}
\mu&=&\frac{1}{\pi}\ln\left[
\sqrt{\left(
\frac{3\lambda}{2\lambda_c}-\frac{1}{2}
\right)^2-1}
-\frac{1}{2}+\frac{3\lambda}{2\lambda_c}
\right]
\\
&=&
\frac{1}{\pi}
\cosh^{-1}\left(
\frac{3\lambda}{2\lambda_c}-\frac{1}{2}
\right)>0
\end{eqnarray}
\end{subequations}
Replacing into~(\ref{eq:small-asymptotics21}) and keeping only the
first two dominant terms (which are of the same order) gives
\begin{equation}
e^{-y_{21}/2}=\frac{-\hr}{3\sqrt{6}\mu i}\left(z-\bar{z}\right)+{\cal O}(\hr^8)
\end{equation}
with
\begin{equation}
z=\left(\frac{\hr}{6\sqrt{3}}\right)^{3i\mu/2}
\left[\frac{\Gamma\left(1-i\frac{\mu}{2}\right)\Gamma(1-i\mu)}{
\Gamma\left(1+i\frac{\mu}{2}\right)\Gamma(1+i\mu)}\right]^{1/2}
\end{equation}
If one chooses $\lambda$ to approach the cut $[\lambda_c,+\infty)$
from above then $\mu$ is changed into $-\mu$ and the final result is
unchanged.

Finally, the potential $y_{21}$ is given by
\begin{equation}
  e^{-y_{21}/2}=\frac{2\hr}{3\mu\sqrt{6}}\sin\left[
  -\frac{3\mu}{2}\ln\frac{\hr}{6\sqrt{3}}-\Psi^{(2:1)}(\mu) \right]
  +{\cal O}(\hr^8)
\end{equation}
where
\begin{equation}
\Psi^{(2:1)}(\mu)=\Imagin\left\{\ln[\Gamma(1-\frac{i\mu}{2})\Gamma(1-i\mu)] 
\right\} 
\end{equation}

For the sake of completeness, we reproduce here small-$r$ asymptotics
for the 1:1 and the 1:2 cases which were computed in
Refs.~\cite{McCoy-Tracy-Wu} and~\cite{TracyWidom-Toda-asympt},
respectively. For the 1:1 electrolyte, when
$\lambda>\lambda_c^{(1:1)}$, the electric potential is given by
\begin{subequations}
\Label{eq:asympt-above11}
\begin{equation}
\Label{eq:asympt-above11-expy}
e^{-y_{11}(\hr)/2}=
\frac{\hr}{4\mu}\sin\left[
-2\mu\ln\frac{\hr}{8}+2 \Psi^{(1:1)}(\mu)
\right]
+{\cal O}(\hr^5)
\end{equation}
with
\begin{equation}
\Label{eq:asympt-above11-mu}
\mu=\frac{1}{\pi}\cosh^{-1}(\pi\lambda)>0
\qquad(1:1)
\end{equation}
and
\begin{equation}
\Label{eq:asympt-above11-psi}
\Psi^{(1:1)}(\mu)=\Imagin\left[\ln\Gamma(1+i\mu)\right]
\end{equation}
\end{subequations}
For the 1:2 electrolyte, when $\lambda>\lambda_c^{(1:2)}$, the asymptotics
are~\cite{TracyWidom-Toda-asympt}
\begin{equation}
e^{-y_{12}(\hr)}=
\frac{\hr}{3\mu\sqrt{3}}\sin\left[-3\mu\ln\frac{\hr}{6\sqrt{3}}
-2\Psi^{(1:2)}(\mu)\right]
+{\cal O}(\hr^4)
\quad(1:2)
\end{equation}
now with
\begin{equation}
\mu=\frac{1}{\pi}
\cosh^{-1}\left(
\frac{1}{2}+\frac{\lambda}{2\lambda_c}
\right)>0
\end{equation}
and
\begin{equation}
\Psi^{(1:2)}(\mu)=\Imagin\left\{
\ln\left[\Gamma\left(\frac{1-i\mu}{2}\right)\Gamma(1-i\mu)\right]
\right\}
\,.
\end{equation}

For practical purposes, since $\mu$ is at most of order $1/|\ln\ha|$,
we can expand the functions
\begin{eqnarray}
  \Psi^{(1:1)}(\mu)&=& -\gamma\mu + {\cal O}(\mu^3)\\
  \Psi^{(1:2)}(\mu)&=& \mu(\frac{3}{2} \gamma +\ln 2)+ {\cal O}(\mu^3)  \\
  \Psi^{(2:1)}(\mu)&=&3\gamma\mu/2 + {\cal O}(\mu^3)
\end{eqnarray}
The asymptotics presented in Eqs.~(\ref{eq:pttedistgrdlambda}) in the
main text use this approximation for $\Psi(\mu)$. 

In principle this means that our expressions in the main text for
$\mu$ are accurate only to order ${\cal O}(1/|\ln \ha|^2)$. However, an explicit
computation using further terms in the Taylor expansion of $\Psi(\mu)$
shows that our results, in the form presented in the main text, are
actually accurate up to order $1/|\ln a|^3$ for $\mu$ and order
$1/|\ln \ha|^4$ for the effective charges.

To illustrate this, consider, for example, the value of $\mu_{\sat}$
for the 1:1 case. Using a Taylor expansion of $\Psi^{(1:1)}(\mu)$ up to
order $\mu^3$ gives
\begin{widetext}
\begin{equation}
  \label{eq:mu-long}
  \mu_{\sat}^{(1:1)}=\frac{\pi}{2\ln\frac{\ha}{8}}
\left[
  1-\frac{\gamma}{\ln\frac{\ha}{8}}
  +\frac{\gamma^2}{\left(\ln\frac{\ha}{8}\right)^2}
  -\frac{\gamma^3-\frac{\pi^2}{24}\psi^{(2)}(1)
  }{\left(\ln\frac{\ha}{8}\right)^3}
  \right]+O\left((\ln\ha)^{-5}\right)
\end{equation}
where $\psi(x)=d\ln\Gamma(x)/dx$ is the digamma function and
$\psi^{(2)}(x)$ its second derivative. Replacing this expression into
Eqs.~(\ref{eq:connection11b}) and~(\ref{eq:xieff-lambda}), yields
the effective charge at saturation
\begin{equation}
  \label{eq:xi-long}
  \xi_{\text{sat}}^{(1:1)}=
  \ha K_1(\ha)\left[
    \frac{2}{\pi}
    +\frac{\pi^3}{4\left(\ln\frac{\ha}{8}\right)^{2}}
    -\frac{\gamma\pi^3}{2\left(\ln\frac{\ha}{8}\right)^{3}}
    +\frac{144\gamma^2\pi^3+\pi^7}{192\left(\ln\frac{\ha}{8}\right)^{4}}
    -\frac{\pi^3(48\gamma^3+\gamma\pi^4+\pi^2\psi^{(2)}(1))/48}{
      \left(\ln\frac{\ha}{8}\right)^{5}}
    +O\left((\ln\ha)^{-6}\right)
    \right]
\end{equation}
\end{widetext}
A direct comparison of Eqs.~(\ref{eq:xi-long}) and~(\ref{eq:mu-long})
with~(\ref{eq:musat11}) and~(\ref{eq:xisat11}) shows that the (more
compact) expressions presented in the main text are indeed accurate up
to order $1/|\ln a|^3$ for $\mu$ and order $1/|\ln \ha|^4$ for
$\xi_{\sat}$.

\end{appendix}


\end{document}